\documentclass[twocolumn,twocolappendix]{aastex701}

\usepackage{newtx}
\DeclareSymbolFont{CMop}{OT1}{cmr}{m}{n}
\SetSymbolFont{CMop}{bold}{OT1}{cmr}{bx}{n}
\DeclareSymbolFont{CMlet}{OML}{cmm}{m}{it}
\SetSymbolFont{CMlet}{bold}{OML}{cmm}{b}{it}
\DeclareSymbolFont{CMSy}{OMS}{cmsy}{m}{n}
\SetSymbolFont{CMSy}{bold}{OMS}{cmsy}{b}{n}
\DeclareSymbolFont{AMSa}{U}{msa}{m}{n}
\SetSymbolFont{AMSa}{bold}{U}{msa}{m}{n}

\AtBeginDocument{%
    \DeclareMathSymbol{/}{\mathord}{CMop}{"2F}%
    \DeclareMathSymbol{+}{\mathbin}{CMop}{"2B}%
    \DeclareMathSymbol{-}{\mathbin}{CMSy}{"00}%
    \DeclareMathSymbol{=}{\mathrel}{CMop}{"3D}%
    \DeclareMathSymbol{<}{\mathrel}{CMlet}{"3C}%
    \DeclareMathSymbol{>}{\mathrel}{CMlet}{"3E}%
    \DeclareMathSymbol{\leqslant}{\mathrel}{AMSa}{"36}%
    \DeclareMathSymbol{\geqslant}{\mathrel}{AMSa}{"3E}%
    \DeclareMathSymbol{\lesssim}{\mathrel}{AMSa}{"2E}%
    \DeclareMathSymbol{\gtrsim}{\mathrel}{AMSa}{"26}%
    \DeclareMathSymbol{\sim}{\mathrel}{CMSy}{"18}%
    \DeclareMathSymbol{\pm}{\mathrel}{CMSy}{"06}%
    \DeclareMathSymbol{\approx}{\mathrel}{CMSy}{"19}%
    \DeclareMathSymbol{\times}{\mathbin}{CMSy}{"02}%
    \DeclareMathSymbol{\Delta}{\mathalpha}{CMop}{1}%
    \DeclareMathSymbol{\Omega}{\mathalpha}{CMop}{10}%
    \DeclareMathSymbol{\alpha}{\mathalpha}{CMlet}{11}%
    \DeclareMathSymbol{\beta}{\mathalpha}{CMlet}{12}%
    \DeclareMathSymbol{\delta}{\mathalpha}{CMlet}{14}%
    \DeclareMathSymbol{\mu}{\mathalpha}{CMlet}{22}%
    \DeclareMathSymbol{\sigma}{\mathalpha}{CMlet}{27}%
}

\received{2025 October 27}
\revised{2026 February 7}
\accepted{2026 February 10}

\shorttitle{Quenching galaxies in TNG50}
\shortauthors{Lawlor-Forsyth et al.}

\begin{document}

\title{Identifying and distinguishing quenching galaxies with spatially resolved star formation in TNG50}

\author[0000-0002-2958-0593]{Cameron~Lawlor-Forsyth}
\affiliation{Department of Physics and Astronomy, University of Waterloo, Waterloo, ON N2L 3G1, Canada}
\affiliation{Waterloo Centre for Astrophysics, University of Waterloo, Waterloo, ON N2L 3G1, Canada}
\email[show]{clawlorforsyth@uwaterloo.ca}

\author[0000-0003-4849-9536]{Michael~L.~Balogh}
\affiliation{Department of Physics and Astronomy, University of Waterloo, Waterloo, ON N2L 3G1, Canada}
\affiliation{Waterloo Centre for Astrophysics, University of Waterloo, Waterloo, ON N2L 3G1, Canada}
\email{mbalogh@uwaterloo.ca}

\author[0000-0001-6245-5121]{Elizaveta~Sazonova}
\affiliation{Department of Physics and Astronomy, University of Waterloo, Waterloo, ON N2L 3G1, Canada}
\affiliation{Waterloo Centre for Astrophysics, University of Waterloo, Waterloo, ON N2L 3G1, Canada}
\email{liza.sazonova@uwaterloo.ca}

\author[0009-0009-2522-3685]{Cameron~R.~Morgan}
\affiliation{Department of Physics and Astronomy, University of Waterloo, Waterloo, ON N2L 3G1, Canada}
\affiliation{Waterloo Centre for Astrophysics, University of Waterloo, Waterloo, ON N2L 3G1, Canada}
\email{crmorgan@uwaterloo.ca}

\author[0000-0003-3255-3139]{Sean~L.~McGee}
\affiliation{School of Physics and Astronomy, University of Birmingham, Birmingham B15 2TT, UK}
\email{smcgee@star.sr.bham.ac.uk}

\author[0000-0001-5851-1856]{Gregory~H.~Rudnick}
\affiliation{Department of Physics and Astronomy, University of Kansas, Lawrence, KS 66045, USA}
\email{grudnick@ku.edu}

\correspondingauthor{C.~Lawlor-Forsyth}

\begin{abstract}
Using the TNG50 simulation, we determine observationally motivated metrics that can distinguish quenching galaxies from star forming galaxies for $M_{*} \geqslant 10^{9.5}~M_{\odot}$, based on the spatial distribution of their stellar populations. Quenching galaxies are not fully quenched but have low levels of ongoing star formation that decreases over time. The morphological metrics consider the concentration of star formation, size of the star forming disk, and characteristic radii that trace sharp truncations of star formation. These metrics can separate simulated quenching galaxies based on morphology into populations where star formation is suppressed inside-out and outside-in. Inside-out quenched galaxies are more likely to be the most massive galaxy within their halo in the field, while outside-in quenched galaxies are satellites residing in dense environments and begin quenching ${\sim} 1~\text{Gyr}$ after being accreted. Outside-in quenched galaxies typically take ${\sim} 1.5~\text{Gyr}$ to quench, and inside-out quenched galaxies can take up to ${\sim} 3.5~\text{Gyr}$, where the duration of quenching is a function of stellar mass. We find that each population of quenched galaxy experiences evolution of their morphological metrics, where the different quenched populations reside in unique locations in parameter space. Galaxies in the later stages of quenching are more easily distinguished than those in the early stages, when compared to star forming galaxies. In addition, inside-out quenched galaxies can be distinguished compared to outside-in quenched galaxies, and the progress through the quenching episode can be estimated for both populations. These results have broad implications for distinguishing quenching galaxies in large galaxy surveys.
\end{abstract}


\section{Introduction}

It has been well known for several decades that galaxies exhibit a strong bimodality in color \citep[e.g.,][]{strateva2001,blanton2003,baldry2004,bell2004,driver2006,wyder2007,brammer2009,gavazzi2010,peng2010,taylor2015}, with a prominent red sequence alongside a blue cloud. Galaxies residing in the red sequence have low rates of star formation \citep[e.g.,][]{blanton2009,bluck2020a,bluck2020b}, and therefore appear red and dead \citep[e.g.,][]{brammer2009,schawinski2014}, while actively star forming galaxies comprise the blue cloud \citep[e.g.,][]{strateva2001,bell2004,bluck2020a,bluck2020b}. Galaxies that have their star formation suppressed or quenched will move from the blue cloud to settle onto the red sequence \citep[e.g.,][]{bell2004,faber2007,martin2007}. The lack of evolution of the massive end of the star forming stellar mass function with redshift is considered strong supporting evidence for quenching \citep[e.g.,][]{faber2007,ilbert2010,ilbert2013,pozzetti2010,muzzin2013,weaver2023}, and there is a corresponding build-up of quiescent galaxies with time \citep[e.g.,][]{drory2009,tomczak2014,davidzon2017}.

Many physical scenarios have been proposed to explain galaxy quenching. Lack of gas accretion or cooling, the inability of cold gas to form stars, the rapid consumption of cold gas, and gas outflow or removal are some of the primary considerations for different quenching mechanisms \citep{man2018}. For example, active galactic nuclei feedback has long been thought to be an additional energy source that will prevent gas cooling \citep[e.g.,][]{dimatteo2005,bower2006,croton2006,mcnamara2007,mcnamara2012,fabian2012}. Similarly, disk or bar instabilities can lead to internal secular quenching \citep{cole2000,martig2009,gensior2020}. Environmental effects must also be considered as an independent pathway for quenching \citep[e.g.,][]{peng2010}. The interaction between a galaxy falling into a large structure like a galaxy cluster and its intracluster medium will lead to ram pressure being exerted on the galaxy \citep{gunn1972,boselli2022,dowling2023}. Strangulation or starvation \citep{larson1980,kawata2008,peng2015} prevents galaxies from replenishing their cold gas supply, while gravitational harassment \citep{moore1996,moore1998,moore1999} is due to close galaxy--galaxy interactions in dense environments like clusters.

While these various mechanisms will all eventually quench a galaxy in the absence of gas inflow, it is expected as an ansatz that these mechanisms will present different morphological spatial signatures \citep{zhang2019} for young stellar populations that trace star formation. However, it is not clear how the signatures express themselves in the full dynamic environment of galaxies. For instance, it has been proposed that active galactic nuclei feedback in the inner regions of a galaxy will lead to a lack of young stars in these central regions \citep[e.g.,][]{dubois2013}. Environmentally, ram pressure stripping will progressively remove gas from the outer regions of a galaxy \citep[e.g.,][]{gunn1972,abadi1999,quilis2000,balogh2000,vollmer2001,mccarthy2008,boselli2008,poggianti2016,poggianti2017}, and it should lead to galaxies with centrally concentrated star formation. Strangulation or starvation is thought to similarly shrink the star forming disk of a galaxy \citep{finn2018,morgan2024}, while gravitational harassment can lead to tidal features \citep{farouki1981} and nuclear starbursts \citep{moore1996}.

Spatially resolved observations are therefore an ideal avenue to best understand the complicated nature of galaxy quenching, given the additional information provided by such observations compared to integrated galaxy properties. Early work utilizing spatially resolved observations focused on color distributions to create maps of relative age, star formation rate, and stellar mass for nearby and intermediate-redshift galaxies \citep[e.g.,][]{abraham1999,bell2000,zheng2004,wu2005,zibetti2009}. Since then, several large integral field unit programs including the Calar Alto Legacy Integral Field Area survey \citep[CALIFA;][]{sanchez2012}, the Sydney-AAO Multi-object Integral field spectrograph galaxy survey \citep[SAMI;][]{croom2012,bryant2015}, and the Mapping Nearby Galaxies at Apache Point Observatory \citep[MaNGA;][]{bundy2015} have made spatially resolved observations available in large quantities for thousands of nearby galaxies \citep{croom2021,abdurrouf2022}. This has enabled substantial progress in studies of active galactic nuclei host galaxies \citep[e.g.,][]{sanchez2018}, age, metallicity, and specific star formation rate gradients \citep[e.g.,][]{gonzalez2014,gonzalez2015,gonzalez2016,belfiore2017b,belfiore2018,goddard2017}, oxygen abundance gradients \citep{sanchez2014,sanchez2016}, the dynamical state--morphology relation \citep[e.g.,][]{cortese2016,brough2017,falcon2017,graham2018}, ionized gas excitation properties \citep{belfiore2016,zhang2017}, the morphology--density relation \citep[e.g.,][]{schaefer2017,schaefer2019,owers2019}, the resolved star forming main sequence \citep[e.g.,][]{cano2016,ellison2018,medling2018}, and stellar mass assembly \citep[e.g.,][]{perez2013,ibarra2016}.

Of particular note is work with integral field unit observations that has been done to understand the nature of inside-out quenching, where star formation is initially suppressed in the central regions of a galaxy and proceeds outward \citep[e.g.,][]{gonzalez2015,gonzalez2016,li2015,tacchella2015,tacchella2016,tacchella2018,belfiore2017a,belfiore2018,goddard2017,ellison2018,liu2018,medling2018,sanchez2018,spindler2018,wang2018,guo2019,lin2019,woo2019,neumann2020,leung2025}. In addition to integral field unit observations, several authors have used high-resolution photometry to understand inside-out quenching through the use of spectral energy distribution fitting \citep[e.g.,][]{morselli2019,nelson2021}. Notably, \citet{morselli2019} find that galaxies that are below the star forming main sequence \citep[e.g.,][]{brinchmann2004} show a clear suppression of star formation in their central regions, and this suppression becomes stronger with increasing distance below the main sequence. Similarly, \citet{nelson2021} find, by comparing their high-resolution observations to simulated galaxies from the high-resolution TNG50 \citep{nelson2019b,pillepich2019} run of IllustrisTNG \citep[e.g.,][]{nelson2018,pillepich2018b,springel2018}, that massive galaxies below the main sequence likewise display centrally suppressed star formation in both observations and simulations.

In addition to the many studies that have investigated the nature of inside-out quenching, others have examined outside-in quenching, where star formation is initially suppressed in the outer regions of a galaxy and proceeds inward \citep[e.g.,][]{schaefer2017,schaefer2019,medling2018,chen2019,owers2019,wang2022,cheng2024}. Considering integral field unit studies and SAMI galaxies, these works have shown that the fraction of star forming galaxies with centrally concentrated star formation increases with increasing environmental density \citep{schaefer2017,wang2022}, and group galaxies display lower levels of specific star formation than field galaxies, along with enhanced centrally concentrated star formation \citep{schaefer2019}. As well, SAMI galaxies in denser environments display suppressed specific star formation radial profiles from the outside in \citep{medling2018}, and galaxies with strong H$\delta$ absorption \citep[indicative of a lack of ongoing star formation;][]{couch1987,poggianti1999} are more often found in clusters than the field. These galaxies have the strongest H$\delta$ absorption in their outskirts \citep{owers2019}. In addition, \citet{finn2018} found smaller star forming disks in cluster galaxies using $24~\mu \text{m}$ Spitzer \citep{werner2004} imaging, while \citet{finn2023} found an excess of star forming galaxies with suppressed star formation rates in clusters. Similar results were found by \citet{morgan2024} using different techniques in Virgo, as well as by \citet{conger2025}. Furthermore, several studies using integral field unit data have investigated ram pressure stripping in galaxies \citep[e.g.,][]{poggianti2017,poggianti2019,poggianti2025,vulcani2020,roberts2022}.

Beyond integral field unit programs, the next decade will see continued growth in large galaxy surveys with current, upcoming, and proposed missions that provide high spatial resolution, using facilities like the James Webb Space Telescope \citep[JWST;][]{gardner2006,gardner2023}, the Nancy Grace Roman Space Telescope \citep[NGRST;][]{spergel2015,akeson2019}, and the Cosmological Advanced Survey Telescope for Optical and UV Research \citep[CASTOR;][]{cote2025}. Given the great diversity and complexity of different quenching mechanisms, similar analyses of simulated data will be essential for interpretation. In recent years, many large-scale cosmological simulations have become available \citep[e.g.,][]{genel2014,vogelsberger2014b,schaye2015,nelson2018,pillepich2018b,springel2018}. 

To that end, we aim to use simulations to identify quenching galaxies using simple, observationally motivated metrics that encode information about the physical nature of quenching, through samples of galaxies that end up quenched. These morphological metrics should easily separate quenching galaxies from star forming galaxies based on the distribution of star formation alone, be capable of separating quenching galaxies based on the spatial signature of quenching (e.g., inside-out versus outside-in), provide information on the length of time that quenching takes, and provide insight into how far along the quenching process a galaxy might be. Using simulated galaxies facilitates tracking the evolution of these quenching systems and thus the evolution of these morphological metrics. With the high spatial resolution provided by recent cosmological simulations, we are in a position to better link our findings to what we can observe, and make predictions for the morphological evolution of quenched galaxies in the real Universe.

This paper is structured as follows: in Section~\ref{sec:methods}, we describe the data coming from the high-resolution TNG50 simulation, our analysis procedure to identify quenched galaxies in the simulation, and our metrics that we use to classify galaxies based on morphological evolution. In Section~\ref{sec:results}, we present our results, differences between quenched populations, show how our metrics evolve with time, and illustrate how machine learning can be used to best separate different populations. In Section~\ref{sec:discussion}, we discuss our results, in particular focusing on the complementary nature of these morphological metrics when coupled with other environmental measures related to phase space, and the feasibility and implications of using such metrics in upcoming large-scale galaxy surveys. Finally, in Section~\ref{sec:summary}, we present our summary and conclusion.

Throughout this paper, we adopt a flat $\Lambda$CDM cosmology that is consistent with the TNG simulation \citep{weinberger2017,pillepich2018a}, based on the Planck intermediate results \citep{planck2016}: $H_{0} = 67.74~\text{km s}^{-1}~\text{Mpc}^{-1}$, $\Omega_{\text{m}} = 0.3089$, $\Omega_{\Lambda} = 0.6911$, and $\Omega_{\text{b}} = 0.0486$. Additionally, we frequently use the term ``morphology'' when referring to the spatial distribution of newly formed stars, in contrast with usage of the term ``galaxy morphology'' in the literature, which generally describes the visual morphology of the total stellar component.

\section{Methodology}\label{sec:methods}

\subsection{IllustrisTNG}

The simulations that comprise the IllustrisTNG suite\footnote{\url{https://www.tng-project.org}} \citep[e.g.,][]{nelson2018,nelson2019a,pillepich2018b,springel2018} are state-of-the-art cosmological gravo-magneto-hydrodynamical simulations and feature an updated version of the galaxy formation model \citep{weinberger2017,pillepich2018a} included in Illustris \citep{vogelsberger2013,torrey2014}. These simulations are the successor to the original Illustris simulation \citep{genel2014,vogelsberger2014a,vogelsberger2014b}, which uses the moving-mesh code \textsc{arepo} \citep{springel2010,pakmor2011,pakmor2016,weinberger2020}. Various considerations were taken into account when developing the IllustrisTNG model, but chief among these are tuning the model to match observations of the global star formation rate (SFR) density across $z = 8$ to $z = 0$, the galaxy mass function, the stellar mass--halo mass relation, the black hole--stellar mass relation, halo gas fractions, and galaxy sizes, all at $z = 0$ \citep{rodriguez2019}. In the IllustrisTNG simulations, subhalos (i.e. galaxies) are identified from their parent halos using the \textsc{subfind} algorithm \citep{springel2001,dolag2009}, while a cosmology consistent with the \citet{planck2016} results is adopted, and stellar populations are drawn from a \citet{chabrier2003} initial mass function \citep{pillepich2018a}. All simulation data from IllustrisTNG are publicly available \citep{nelson2019a}.

In the present work, we use the highest-resolution simulation in the suite, TNG50-1 (hereafter referred to as TNG50), which encompasses a cubic volume of $35h^{-1} \approx 51.7~\text{Mpc}$ on a side \citep{nelson2019b,pillepich2019}, and follows the evolution of $2 \times 2160^{3}$ resolution elements (dark matter and baryons), where the average mass of a baryonic resolution element is $8.5 \times 10^{4}~M_{\odot}$. The spatial scale of the simulation is essentially set by the gravitational softening length of the dark matter and stellar particles, which is 288 pc at $z = 0$, while the gas softening is adaptive with a minimum of 74 pc \citep[comoving;][]{nelson2019b,pillepich2019}. The average cell size of star forming gas in the interstellar medium of galaxies is ${\sim}$100--140 pc \citep{nelson2019b,pillepich2019}. TNG50 has better mass resolution by more than an order of magnitude over the flagship TNG100-1, with two and a half times better spatial resolution \citep{nelson2019b}, placing TNG50 near modern `zoom' simulations of individual galaxies in terms of mass resolution, but in a large cosmological volume.

\subsection{Distinguishing star forming galaxies from quenched galaxies using star formation histories}\label{subsec:quenched_sample}

\subsubsection{Selecting a galaxy sample}\label{subsubsec:sample}

We begin our analysis by considering fundamental properties of all subhalos (galaxies) present in TNG50 at $z = 0$, including stellar masses and instantaneous star formation rates (SFRs). We limit our analysis to galaxies with a total stellar mass of at least $M_{*} = 10^{9.5}~M_{\odot}$ by $z = 0$, to ensure that specific SFR ($\text{sSFR} = \text{SFR}/M_{\odot}$) radial profiles extend to a sufficient distance with adequate sampling (i.e. we essentially require a sufficient number of stellar particles per galaxy; see Section~\ref{subsec:metrics}). We find 1666 total galaxies with $M_{*} \geqslant 10^{9.5}~M_{\odot}$ at $z = 0$ which comprise our fiducial sample. We additionally consider masses for the dark matter halos in which the subhalos are embedded, in order to describe their environment.

We consider the merger history and evolution for each galaxy, and track them back through ``main progenitor branch'' files that encode fundamental information about a galaxy as a function of time through the simulation, similar to what was previously used for all subhalos at $z = 0$. As these files describe the merger history of each galaxy, they link a given galaxy at an arbitrary snapshot with the corresponding progenitor galaxy at the proceeding snapshot, as well as the descendant galaxy at the subsequent snapshot \citep{nelson2015}. As galaxies will experience many mergers, by considering only the most massive galaxy involved in a merger, the main branch along the progenitor tree can be tracked \citep[e.g.,][]{genel2018,nelson2019a}. Within these main progenitor branch files, we have access to additional information such as the stellar mass history, which describes the stellar mass of the galaxy as a function of time, the instantaneous star formation history, which describes the instantaneous SFR of the galaxy as a function of time, as well as the position of the galaxy within the simulation volume, once again as a function of time. We additionally obtain the radius that encompasses half the stellar mass (which functions as the effective radius, $R_{\text{e}}$), for the simulated galaxies as a function of time.

For each snapshot along the main progenitor branch for each galaxy, we use three-dimensional ``cutouts'' of the simulation volume, which encapsulates every bound particle to a given galaxy at that snapshot. We focus on parameters related to the stellar particles (i.e. position, initial birth mass, current mass, and formation time).

\subsubsection{Determining star formation histories and the main sequence}\label{subsubsec:sfms}

We subsequently determine time-averaged total star formation histories (SFHs) for each galaxy in our fiducial sample by employing the results of \citet{donnari2019} and \citet{pillepich2019}, who produced catalogs\footnote{\url{https://www.tng-project.org/data/docs/specifications/\#sec5b}} of SFRs across various time intervals (e.g., 10, 50, 100, 200, and 1000~Myr) for the simulations in the IllustrisTNG suite. In their work, \citet{donnari2019} and \citet{pillepich2019} computed SFRs across the aforementioned timescales by considering all stellar particles formed within the given amounts of time prior to each snapshot (in various physical apertures) through summation of their initial (birth) masses. We use total SFRs, which are computed by including all gravitationally bound stellar particles of a given subhalo, and select $100~\text{Myr}$ as our nominal duration for our time-averaged SFRs. We choose $100~\text{Myr}$, as it reflects the upper limit of a timescale that is often thought to describe ``recent'' star formation for an entire galaxy \citep{calzetti2013} and is the timescale best traced with ultraviolet and infrared observations \citep[e.g.,][]{kennicutt1998b,madau2014}. These time-averaged SFRs are in contrast to the instantaneous SFRs tabulated within the main progenitor branches, which are not observationally motivated. By working through the main progenitor branch for each galaxy and tracking the time-averaged SFRs, we create time-averaged SFHs. When we subsequently refer to the SFR of a galaxy or to attributes or properties related to star formation, we are implicitly using time-averaged values over the above described $100~\text{Myr}$ interval.

\begin{figure}[t]
    \centering
    \includegraphics[width=\columnwidth]{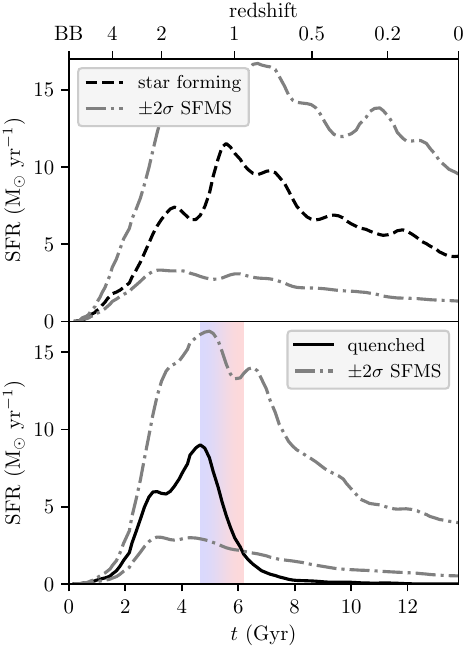}
    \caption{Top: An SFH for a normal star forming galaxy ($M_{*} = 10^{10.72}~M_{\odot}$ at $z = 0$; dashed black line), shown as a function of time (age of the Universe). Additionally shown are the ${\pm} 2 \sigma$ limits from the median SFH of SFMS galaxies in a similar stellar mass range (${\pm} 0.1~\text{dex}$) at every snapshot (dotted--dashed gray lines). Due to the slightly stochastic nature of SFHs over short timescales, we opt to smooth our SFHs with a Gaussian filter, where the smoothing kernel is one snapshot. Bottom: An SFH for a quenched galaxy ($M_{*} = 10^{10.21}~M_{\odot}$ at $z = 0$; solid black line), where the primary quenching event is highlighted using a vertical colored band. Also shown are the corresponding percentiles (gray dotted--dashed lines) for SFMS galaxies with a similar stellar mass (${\pm} 0.1~\text{dex}$) at every snapshot. Once the SFH for the displayed quenched galaxy drops below the 2.5th percentile, it remains below this threshold, thereby defining this galaxy as quenched. The primary quenching episode for this galaxy lasts ${<} 2~\text{Gyr}$, and by $z = 0$ the galaxy has been quenched for ${>} 7~\text{Gyr}$. The subhalo ID for this quenched galaxy is 198186 at $z = 0$.}
    \label{fig:SFHs}
\end{figure}

While the SFHs are being computed for every galaxy in the sample, we simultaneously determine the star forming main sequence \citep[SFMS, or galaxy main sequence; e.g.,][]{brinchmann2004,daddi2007,elbaz2007,noeske2007,peng2010,rodighiero2011,speagle2014,davies2019,bluck2022}. We adopt the ``ridge line method'' \citep[e.g.,][]{pillepich2019,donnari2021a,donnari2021b,bottrell2024} described in \citet{donnari2019} to determine the SFMS, where galaxies are first binned by snapshot stellar mass and the median (log) SFR is determined for galaxies within a given bin. We use a bin size of $0.2~\text{dex}$, and only consider galaxies with $M_{*} \leqslant 10^{10.5}~M_{\odot}$, given that this mass scale roughly corresponds with the location where the SFMS deviates from linearity as determined observationally \citep[e.g.,][]{whiteaker2012,whitaker2014} and in TNG \citep{donnari2021b}. Galaxies with SFRs that are far (${>} 1~\text{dex}$) from the current estimate of the median SFR within the bin are subsequently removed from further analysis, and the median is recomputed. This procedure is repeated iteratively until the median SFR converges and no further iterations are required. These final median SFRs per bin are used with the bin centers to fit a linear function of the form \citep{donnari2019}
\begin{equation*}
    \langle \log{(\text{SFR}/M_{\odot}~\text{yr}^{-1})} \rangle = \alpha(z) \log{(M_{*}/M_{\odot})} + \beta(z),
\end{equation*}
where $\langle \log{(\text{SFR})} \rangle$ are the median (log) SFRs, and $\alpha(z)$ and $\beta(z)$ are the slope and intercept of the resulting fits, respectively, as a function of redshift. For $M_{*} > 10^{10.5}~M_{\odot}$, the SFMS is then linearly extrapolated using the fitted values. We take SFMS galaxies to be those with $\Delta \log{(\text{SFR})} \geqslant -0.5$, consistent with \citet{pillepich2019}. This procedure is then repeated for every snapshot to determine the SFMS as a function of time.

\begin{figure}
    \centering
    \includegraphics[width=\columnwidth]{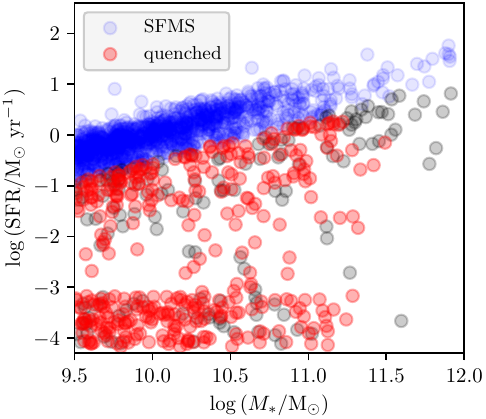}
    \caption{Star formation rate as a function of stellar mass for the fiducial sample (1666 galaxies with $M_{*} \geqslant 10^{9.5}~M_{\odot}$; see Section~\ref{subsubsec:sample}) at $z = 0$. Galaxies that populate the SFMS are shown in blue. Galaxies with low levels of star formation ($\text{SFR} \approx 0$) have their SFR set to a small, random nonzero value between ${\sim} 10^{-3}~M_{\odot}~\text{yr}^{-1}$ and ${\sim} 10^{-4}~M_{\odot}~\text{yr}^{-1}$ for visualization purposes. Galaxies that satisfy our quenched definition (see Section~\ref{subsubsec:quenched}) are shown in red and do not necessarily have $\text{SFR} = 0$, while galaxies that do not populate the SFMS (as defined in Section~\ref{subsubsec:sfms}) and that are not quenched are shown in black. These galaxies reside below the SFMS.}
    \label{fig:SFMS}
\end{figure}

An example SFH for a normal star forming galaxy can be seen in the top panel of Figure~\ref{fig:SFHs}, along with lines denoting the ${\pm} 2 \sigma$ spread of SFHs for comparison SFMS galaxies that have a similar stellar mass at every snapshot. In addition, we show a plot illustrating the SFMS for $z = 0$ in Figure~\ref{fig:SFMS}.

\subsubsection{Identifying quenched galaxies}\label{subsubsec:quenched}

With the star forming main sequence as a function of time determined, we now aim to find a sample of quenched galaxies at $z = 0$. This sample will act as a sample of quenching galaxies at earlier times. We begin with all galaxies not part of the SFMS at $z = 0$ and consider their SFHs as compared to main-sequence galaxies in a similar mass range (${\pm} 0.1~\text{dex}$, though this range can grow if necessary; see below). We impose an additional requirement that each potential quenching galaxy be compared to at least 50 main-sequence galaxies, in order to reduce the effect of statistical uncertainties. If a given galaxy does not have 50 comparison SFMS galaxies within ${\pm} 0.1~\text{dex}$, the stellar mass bin is increased in increments of ${\pm} 0.005~\text{dex}$, until 50 comparison main-sequence galaxies have been found. This requirement is typically only necessary for the most massive galaxies ($M_{*} > 10^{11}~M_{\odot}$ at $z = 0$), and even then it only increases the size of the stellar mass bin modestly, usually much less than ${\pm} 0.05~\text{dex}$.

For each galaxy, we compare its SFH to the SFHs of its comparison main-sequence galaxies (having a similar stellar mass at each snapshot). Due to the stochastic nature of SFHs in TNG over short timescales, we choose to smooth the selected SFH using a Gaussian filter (smoothing kernel, $\sigma$, equal to one snapshot, which is similar to a moving average) to facilitate simpler comparisons with the main-sequence comparison population. Our selected smoothing kernel ensures that broad-scale features remain intact while minimizing large stochastic fluctuations from snapshot to snapshot, especially those where there is a single brief increase (or decrease) in star formation, before returning to the prior level. This smoothing is qualitatively similar to using a slightly longer timescale to measure star formation, but still ${<} 200~\text{Myr}$. Galaxies are considered quenched if
\begin{itemize}
    \item the galaxy's SFH is above $-2 \sigma$ of the median SFH for the comparison SFMS galaxies for all times prior to falling $2 \sigma$ below that median and
    \item the galaxy's SFH subsequently stays at least $2 \sigma$ below that median SFH for all subsequent times.
\end{itemize}
This definition is illustrated in the bottom panel of Figure~\ref{fig:SFHs}, where the SFH for the considered galaxy quenches after about six billion years and remains quenched thereafter. This conservative definition of a quenched galaxy ensures that we arrive at a sample that has a single quenching episode for each quenched galaxy, which facilitates a straightforward consideration of the evolution of their quenching attributes (see Section~\ref{subsec:signatures}). This definition also minimizes ``contamination'' in the overall quenching signal from periods of increasing star formation. Therefore, galaxies that satisfy the first criterion above, but which rejuvenate (and subsequently do not satisfy the second criterion) are excluded from our sample.

Using this definition, out of the total 1666 galaxies with $M_{*} > 10^{9.5}~M_{\odot}$ by $z = 0$, we find 361 quenched galaxies, 1160 star forming main-sequence galaxies, and 145 other galaxies below the SFMS (which do not fit our definition of ``quenched''; see above). In Figure~\ref{fig:SFMS}, we show the quenched galaxies with red, star forming main-sequence galaxies with blue, and the other galaxies below the star forming main sequence in black. Of the 145 galaxies below the SFMS at $z = 0$, 39 have no star formation, while the remaining 106 have nonzero SFRs, and the majority (109) of both of these populations are comprised of galaxies that experienced rejuvenation at some point. Given our definition above, quenched galaxies will not necessarily have $\text{SFR} = 0$, as evidenced by the red points in Figure~\ref{fig:SFMS} that have $10^{-3} \lesssim \text{SFR}/M_{\odot}~\text{yr}^{-1} \lesssim 10^{-1}$. Henceforth, we refer to galaxies which have completed quenching as ``quenched'' (generally when discussing populations at $z = 0$), while we use the term ``quenching'' to refer to systems that are actively undergoing the quenching process.

\begin{figure}
    \centering
    \includegraphics[width=\columnwidth]{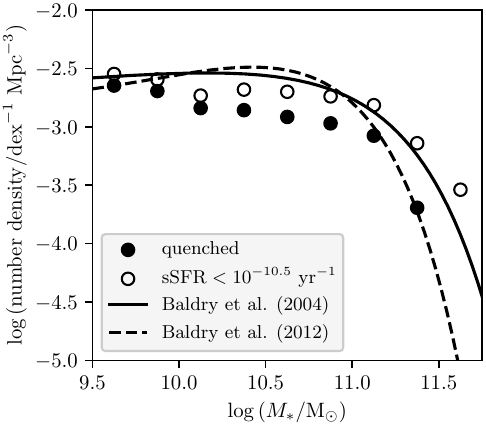}
    \caption{The galaxy stellar mass function of our 361 quenched galaxies (black dots; $z = 0$), shown alongside two fitted galaxy stellar mass functions for local ($0.004 < z < 0.08$) red galaxies \citep[dashed line;][their Figure~10]{baldry2004} in SDSS \citep{york2000}, and local ($z < 0.06$) red galaxies \citep[solid line;][their Figure~15]{baldry2012} in GAMA \citep{driver2011}. \citet{baldry2004} and \citet{baldry2012} define red galaxies based on location in the \textit{u}---\textit{r} color--magnitude diagram. We additionally show a stellar mass function for TNG50 galaxies based on a cut in sSFR using $\log{(\text{sSFR}/\text{yr}^{-1})} < -10.5$ (open circles).}
    \label{fig:SMF}
\end{figure}

We show the derived galaxy stellar mass function for our quenched sample in Figure~\ref{fig:SMF} (black points), alongside observed galaxy stellar mass functions \citep{baldry2004,baldry2012} for local red galaxies in the Sloan Digital Sky Survey \citep[SDSS;][]{york2000} and local red galaxies in the Galaxy And Mass Assembly survey \citep[GAMA;][]{driver2011}. We find that our quenched sample is a subset of all red or low-SFR galaxies, due to our specific and rather particular definition of what constitutes a quenched galaxy, where our definition is based on a galaxy's SFH in a manner that cannot be reproduced from observations. In Figure~\ref{fig:SMF}, we similarly show the stellar mass function for passive or quiescent galaxies selected based on a cut in sSFR, where these galaxies have $\log{(\text{sSFR}/\text{yr}^{-1})} < -10.5$ (open circles). This broader definition for quiescence is a closer match to the field mass functions of red galaxies \citep[e.g.,][]{baldry2004}, and it again highlights how our definition produces a subset of all quiescent galaxies. As above, galaxies that quench, rejuvenate, and quench again are excluded from our sample.

We additionally define the onset of the primary quenching episode to be the most recent significant peak of star formation (i.e. a local maximum) prior to the galaxy falling $2 \sigma$ below its comparison SFMS galaxies. We define a significant peak as any peak that has an SFR of at least 40\% of the maximum SFR throughout the entire SFH. In the bottom panel of Figure~\ref{fig:SFHs}, the onset of quenching is then the absolute maximum of the SFR across the SFH, and is denoted with the start of the colored vertical band. We find that varying the significance threshold from 40\% to 30\% or even 50\% changes the sample size by at most one galaxy, with slight variations in the extent of the quenching episode on a per-galaxy basis, but does not qualitatively affect our main conclusions. When considering the quenching episode duration across all quenched galaxies, changing the significance threshold by ${\pm}$10\% affects the median quenching duration by ${\sim} 200~\text{Myr}$, which is comparable to the length of time between successive snapshots. In addition, for 313 of the quenched galaxies, the onset of quenching began after redshift $z = 1.5$, while only 44 galaxies began quenching during ``cosmic noon'' \citep[$1.5 \lesssim z \lesssim 2.5$;][]{madau2014,forster2020}. Given this, nearly 87\% of the sample has begun quenching in the last ${\sim} 9.5~\text{Gyr}$ of cosmic time. For a small subset (${\sim}$8\%) of the quenched sample, we find that, after the peak in star formation, certain galaxies being quenching, reach a plateau in their level of star formation, and then begin quenching again, though a steady decline in star formation is overwhelmingly the more common trend. Beyond our chosen definition for the onset of quenching, other considerations include when a galaxy crosses a given sSFR threshold, when a galaxy falls off the main sequence, or when a galaxy is a certain distance away from the main sequence. In testing, we found that our median sSFR value at onset is ${\sim} 10 ^{-9.5}~\text{yr}^{-1}$, and so choosing a threshold value similar to this produces onset times that closely align with our preferred onset times on average (${\pm} 200~\text{Myr}$). As well, we determined that, if we elect to use the time when a galaxy falls off the main sequence, this corresponds to ${\sim} 130~\text{Myr}$ after our preferred onset times (on average), which is comparable to the time difference between snapshots. Increasing the required distance from the SFMS works to further shorten the length of the quenching episodes. We subsequently define the end of the primary quenching episode as the snapshot when the galaxy first falls $2 \sigma$ below the median SFR of its comparison SFMS galaxies, which can be seen as the end of the colored vertical band in Figure~\ref{fig:SFHs}. The duration of the quenching episode is given as $\Delta t_{\text{quench}} = t_{\text{term}} - t_{\text{onset}}$.

In addition to the quenched sample, we also create a control sample of normal star forming galaxies. Control galaxies are chosen to have a similar stellar mass ($\Delta \log{[M_{*}/M_{\odot}]} \leqslant 0.1~\text{dex}$) to each quenched galaxy, but with the condition that they reside on the SFMS during the entirety of the quenched galaxy's quenching episode. In general, each quenched galaxy has many corresponding control star forming galaxies, and therefore the control sample of normal star forming galaxies is much larger than the quenched sample, where 1607 of the 1666 galaxies in the fiducial sample were considered part of the control sample at some point. We note that this sample is distinct from the comparison SFMS galaxies that were used to define quenched galaxies earlier, as those galaxies must only reside on the main sequence for a given single snapshot to facilitate comparison, while these control star forming galaxies must be on the main sequence throughout the quenched galaxy's entire quenching episode.

\subsection{Morphological metrics}\label{subsec:metrics}

\begin{figure*}[t]
    \centering
    \includegraphics[width=\textwidth]{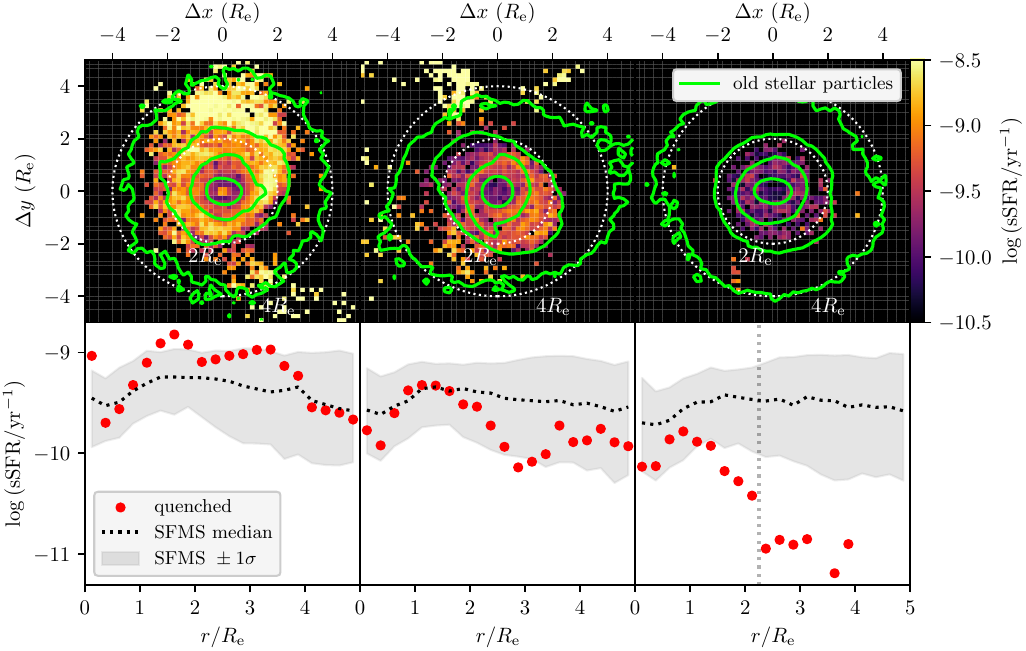}
    \caption{Top: projections of an $M_{*} = 10^{10.21}~M_{\odot}$ (at $z = 0$) quenched galaxy (the same galaxy as shown in the bottom panel of Figure~\ref{fig:SFHs}) at three points during its primary quenching episode. From left to right, the time series shows the snapshot at the onset of quenching, ${\sim}$38\% through the quenching episode, and ${\sim}$75\% through the quenching episode. The postage stamp cutout images show spatial sSFR values, while the green contours trace the stellar mass distribution, and the white dotted circles describe multiples of the effective radius $R_{\text{e}}$. The time from the leftmost panel to the middle panel is $615~\text{Myr}$, while the rightmost panel is an additional $642~\text{Myr}$ after the middle panel. These projections highlight the evolution of the size of the star forming disk relative to the stellar disk, as the star forming disk shows significant outer truncation with increasing time, while the stellar disk remains largely unchanged, in fact shrinking by ${\sim}$10\%. These panels also display the concentration of star formation. Bottom: sSFR radial profiles for snapshots that correspond to the top panels (red points; bin width of $0.25~R_{\text{e}}$). The median sSFR profile for SFMS galaxies at a similar stellar mass is shown with a dotted black line, with the ${\pm} 1 \sigma$ interval shown as a light gray band. The rightmost panel highlights $R_{\text{outer}}$ (vertical dotted line), as there is a clear discontinuity in the sSFR profile for this quenched galaxy. In this case, $R_{\text{outer}}/R_{\text{e}} \approx 2.3$.}
    \label{fig:postage_stamps_and_profiles}
\end{figure*}

Before defining metrics that can be of use in observational surveys, we consider 2D histogram ``postage stamp'' cutouts (based on the distribution of stellar particles) and sSFR radial profiles of quenched galaxies to qualitatively understand morphological features of quenching. Our postage stamp cutouts are projected along the simulation box \textit{xy}-coordinate axes, thereby acting as random angle projections. In Figure~\ref{fig:postage_stamps_and_profiles}, we show a time series ($\Delta t = 1.26~\text{Gyr}$, total) for an example quenched galaxy, which is the same quenched galaxy as in the bottom panel of Figure~\ref{fig:SFHs}. In the top panels of Figure~\ref{fig:postage_stamps_and_profiles}, we see the young ``star forming'' stellar particles (with formation ages ${\leqslant} 100~\text{Myr}$; see Section~\ref{subsubsec:sfms}) shown in the colored histograms, while the stellar disk is shown with contours. Moving through the time series from left to right, we see that the size of the star forming disk evolves relative to the size of the stellar disk, which remains relatively constant, a feature also seen observationally in \citet{zhang2021}. In the bottom panels of Figure~\ref{fig:postage_stamps_and_profiles}, we show sSFR radial profiles for the same snapshots as the upper panels, where the example galaxy's profile is shown in red points (bin width of $0.25~R_{\text{e}}$ out to $5~R_{\text{e}}$), while the median ${\pm} 1\sigma$ band for its comparison SFMS galaxies is shown in gray. We find that this profile also evolves through the quenching episode, and at late times (rightmost panel), this galaxy's sSFR radial profile has a sharp truncation near $2.3~R_{\text{e}}$.

Given the features that we have identified from Figure~\ref{fig:postage_stamps_and_profiles} and from inspecting similar figures for the quenched sample (see Appendix~\ref{app:io_comprehensive_and_evo}), we have defined four metrics that encode information about the distribution of star formation within the quenched galaxies during quenching.

The first metric, $C_{\text{SF}}$, describes the concentration of star formation:
\begin{equation}
    C_{\text{SF}} = \text{SFR}_{{<} 1~\text{kpc}}/\text{SFR}_{\text{total}},
\end{equation}
where $\text{SFR}_{{<} 1~\text{kpc}}$ is the star formation rate in the inner (projected) kiloparsec of the galaxy, and $\text{SFR}_{\text{total}}$ is the star formation rate over the entirety of the galaxy. The center of the galaxy is taken from the main progenitor branch files and is the stellar center. The second metric, $R_{\text{SF}}$, describes the size of the star forming disk relative to the stellar disk:
\begin{equation}
    R_{\text{SF}} = \log{(R_{\text{e, SF}}/R_{\text{e}})},
\end{equation}
where the half-mass radius of the young, star forming particles is $R_{\text{e, SF}}$, while the half-mass radius (or effective radius) of the stellar disk is $R_{\text{e}}$. These radii are again taken relative to the stellar center. The third metric, $R_{\text{inner}}$, describes the inner truncation radius of the galaxy, where the sSFR radial profile of the galaxy has a sharp change in slope:
\begin{equation}\label{eq:Rinner}
    R_{\text{inner}} = \log[\text{sSFR}(r)] \leqslant -10.5 \cap \mathrm{d} (\text{sSFR})/\mathrm{d}r \geqslant 1,
\end{equation}
where $\text{sSFR}(r)$ is the sSFR radial profile, $\mathrm{d} (\text{sSFR})/\mathrm{d}r$ is its first derivative, and $R_{\text{inner}}$ is the outermost radius that satisfies both conditions. For robustness across our considered stellar mass range, we quote $R_{\text{inner}}$ values in relation to $R_{\text{e}}$. We select a threshold sSFR value of $10^{-10.5}~\text{yr}^{-1}$, along with a slope of $+1$, as it was found that these values easily identify the inner truncation radius during testing. We confirm that small changes to the values of the threshold sSFR and slope do not significantly affect our results. $R_{\text{inner}}$ is found by considering the sSFR radial profile and its first derivative, moving outward from $0~R_{\text{e}}$ to $5~R_{\text{e}}$, which we use as the maximum extent for the profiles. For $R_{\text{inner}}$, if no radius is found that satisfies Equation~(\ref{eq:Rinner}), then $0~R_{\text{e}}$ is used as a lower limit. An illustrative example highlighting $R_{\text{inner}}$ is shown in Appendix~\ref{app:io_comprehensive_and_evo}. The fourth and final metric is similar to the inner truncation radius, but is the outer truncation radius, $R_{\text{outer}}$, defined as
\begin{equation}\label{eq:Router}
    R_{\text{outer}} = \log[\text{sSFR}(r)] \leqslant -10.5 \cap \mathrm{d} (\text{sSFR})/\mathrm{d}r \leqslant -1.
\end{equation}
This is the innermost radius that satisfies both conditions. The outer truncation radius is similarly measured moving outward from $0~R_{\text{e}}$ along the sSFR and $\mathrm{d} (\text{sSFR})/\mathrm{d}r$ radial profiles. We likewise quote $R_{\text{outer}}$ values in relation to $R_{\text{e}}$. The upper limit for $R_{\text{outer}}$ is taken to be $5~R_{\text{e}}$ in the event that no radius is found to satisfy Equation~(\ref{eq:Router}). We similarly confirm that small changes to the threshold sSFR and slope do not significantly affect our results. We highlight $R_{\text{outer}}$ in the final (rightmost) panel of Figure~\ref{fig:postage_stamps_and_profiles} for illustration. Finally, as our metrics are defined using two-dimensional projected distances, we confirm that including the full three-dimensional positional information does not significantly change our results. We see only modest changes for $C_{\text{SF}}$ and $R_{\text{inner}}$ (at ${\lesssim} 1~R_{\text{e}}$), while the remaining metrics do not have increased precision, but instead show a negligible scatter around their two-dimensional values (see Appendix~\ref{app:3d_metric_comparison}).

\subsection{Classifying galaxies according to morphological evolution during quenching}\label{subsec:signatures}

\begin{figure}
    \centering
    \includegraphics[width=\columnwidth]{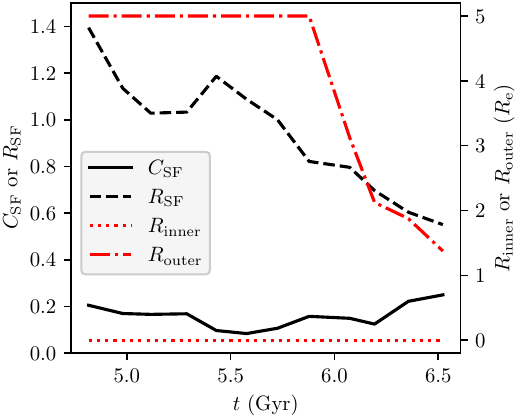}
    \caption{Evolution of the morphological metrics for the same galaxy as in the bottom panel of Figure~\ref{fig:SFHs} and in Figure~\ref{fig:postage_stamps_and_profiles}, where the vertical axis on the left records the values for $C_{\text{SF}}$ (solid black line) and $R_{\text{SF}}$ (dashed black line), while the vertical axis on the right records the values for $R_{\text{inner}}/R_{\text{e}}$ (dotted red line) and $R_{\text{outer}}/R_{\text{e}}$ (dotted--dashed red line). The extent of the \textit{x}-axis reflects the primary quenching episode for this galaxy, which lasts ${<} 2~\text{Gyr}$, and is the same as the colored band seen in the bottom panel of Figure~\ref{fig:SFHs}. This galaxy's morphological evolution is consistent with an outside-in signature, seen through the decrease of $R_{\text{SF}}$ and $R_{\text{outer}}$, along with $C_{\text{SF}}$ modestly increasing. This galaxy shows no evolution for $R_{\text{inner}}$, a feature commonly noticed for galaxies in the outside-in class in our sample. $R_{\text{e}}$ shrinks by ${\sim}$10\% for this galaxy through the quenching episode.}
    \label{fig:metric_example_evolution}
\end{figure}

With the definitions of our metrics in place, we now determine their values as a function of time for all the quenched galaxies in our sample. We focus on the primary quenching episode, by considering the state of these metrics at the onset of quenching ($t_{\text{onset}}$) until the galaxy has dropped $2 \sigma$ below the median for the comparison SFMS galaxies, which we describe as the termination of the primary quenching episode ($t_{\text{term}}$). To illustrate this, we show the evolution of the metrics in Figure~\ref{fig:metric_example_evolution} for the same galaxy as was shown in the bottom panel of Figure~\ref{fig:SFHs}, as well as in Figure~\ref{fig:postage_stamps_and_profiles}.

In Figure~\ref{fig:metric_example_evolution}, by considering the left vertical axis, we see that $R_{\text{SF}}$ (black dashed line) decreases through the quenching episode, while $C_{\text{SF}}$ (solid black line) increases modestly. Additionally, by considering the right vertical axis, we find that $R_{\text{inner}}$ stays fixed at low values close to $0~R_{\text{e}}$ with no evolution, while $R_{\text{outer}}$ evolves strongly, sharply decreasing from $t = 5.8~\text{Gyr}$ to $t = 6.2~\text{Gyr}$.

We create plots such as Figure~\ref{fig:metric_example_evolution} for every quenched galaxy in the sample, and together with plots such as Figure~\ref{fig:postage_stamps_and_profiles}, manually categorize our quenched galaxies into three broad classes. The first contains those that show a morphological evolution that proceeds from the inside out, such that star formation is suppressed in the center of the galaxy and proceeds outward. For this class, $C_{\text{SF}}$ will decrease through the quenching episode, $R_{\text{SF}}$ increases, and $R_{\text{inner}}$ will evolve to higher values of $R_{\text{e}}$ while $R_{\text{outer}}$ stays relatively fixed with no strong evolution. We show corresponding figures similar to Figures~\ref{fig:postage_stamps_and_profiles} and \ref{fig:metric_example_evolution} for an example inside-out quenched galaxy in Appendix~\ref{app:io_comprehensive_and_evo}. The second class is comprised of galaxies whose morphological evolution proceeds from the outside in, where star formation is suppressed in the outer regions of the galaxy and proceeds inward. The galaxy shown in the bottom panel of Figure~\ref{fig:SFHs} and in Figures~\ref{fig:postage_stamps_and_profiles} and \ref{fig:metric_example_evolution} is in this second class. These galaxies generally have increasing $C_{\text{SF}}$ through the quenching episode with decreasing $R_{\text{SF}}$, with relatively constant $R_{\text{inner}}$ along with strongly evolving $R_{\text{outer}}$. The last class of quenched galaxies are those whose morphological evolution does not clearly match either of the first two classes, and is therefore ambiguous, with aspects that are similar to the inside-out class, but with other aspects that more closely resemble the outside-in class. These galaxies often have modestly increasing $C_{\text{SF}}$ through the quenching episode, with relatively constant $R_{\text{SF}}$, with evolution of both $R_{\text{inner}}$ and $R_{\text{outer}}$.

In this classification process, there are situations where the above evolutionary trends are not strictly met for some galaxies, especially when considering the inside-out and outside-in populations. An example is an outside-in quenched galaxy where the evolution of $R_{\text{SF}}$ is modest and does not decrease strongly. For these situations, we consider the evolution of the other morphological metrics, in combination with insights gained through visual inspection of the projected postage stamps in order to classify such a galaxy. Regardless, for the majority of the quenched galaxies (${>}$72\%), we classify them into either the inside-out or outside-in populations without major difficulty. As well, we initially considered an additional class of quenched galaxy driven by starvation$/$uniform suppression, where the star formation is uniformly suppressed throughout the galaxy, but we find no conclusive evidence for this additional class based on our metrics. For a galaxy experiencing uniform suppression, we expect that $C_{\text{SF}}$ would stay relatively constant through the quenching episode, if the assumption of an unchanging star formation radial profile gradient is valid. This should likewise correspond to a relatively constant $R_{\text{SF}}$, and little evolution for $R_{\text{inner}}$ and $R_{\text{outer}}$. However, this type of uniform suppression would become noticeable once the star formation levels approach the detection limit, where evolution could then be seen.

Through this classification process, we find 78 galaxies have an inside-out morphological evolution signature, while 185 have an outside-in signature. The remaining 98 galaxies have a morphological evolution that is ambiguous.

\section{Results}\label{sec:results}

\subsection{Differences between populations}\label{subsec:io_vs_oi}

We investigate differences between our quenched populations by considering their environments, and the relation between when a galaxy was accreted as a satellite and the onset of quenching.

\begin{figure*}[ht]
    \centering
    \includegraphics[width=\textwidth]{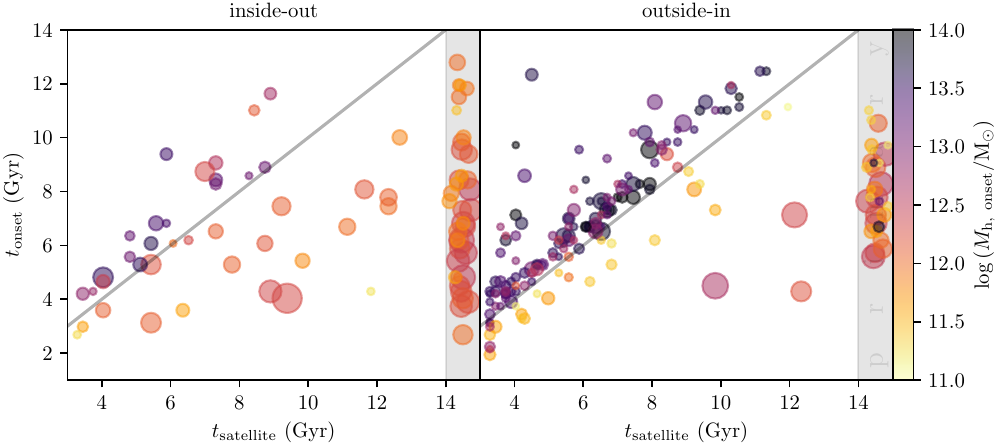}
    \caption{The time corresponding to the onset of quenching for our quenched sample, plotted as a function of the time when the galaxy was accreted as a satellite onto a more massive structure, for two of our quenched populations. We plot a line of equality in gray. For galaxies that have always been primary galaxies, by definition the satellite time is set as the current age of the Universe, and we plot these in a separate region with minor perturbations beyond the current age of the Universe for visual clarity (shaded gray region). We separate the inside-out population (left) from the outside-in population (right). We scale our data points by $z = 0$ stellar mass, where more massive galaxies are larger, and we additionally color-code data points by their host halo mass at the onset of quenching. Halo masses have a range from galaxy-scale halos (lighter colors) to cluster$/$cluster-like mass halos (darker colors). Galaxies below the line of equality have a primary quenching episode that begins before they are accreted as a satellite, while those that sit above equality experience a delay before quenching begins. Galaxies lying along or close to the line of equality have a strong correlation between when they were first accreted and when their primary quenching episode began (e.g., the outside-in population).}
    \label{fig:satellite_time}
\end{figure*}

We obtain $z = 0$ primary$/$satellite flags for each galaxy, which indicate if a given galaxy is the primary galaxy in a given halo. These primary flags are set with the \textsc{subfind} \citep{springel2001,dolag2009} algorithm, and in most cases report the most massive galaxy in the halo. The environments at quenching onset and at $z = 0$ are then determined for every galaxy, based on the mass of the halo that they reside in. At $z = 0$, we define cluster-like environments to be those that have $M_{\text{h}}/M_{\odot} \geqslant 10^{14}$, while high-mass groups have halos with $10^{13.5} < M_{\text{h}}/M_{\odot} < 10^{14}$, low-mass groups have $10^{13} < M_{\text{h}}/M_{\odot} < 10^{13.5}$, and finally galaxy-scale (i.e. field) halos have $M_{\text{h}}/M_{\odot} < 10^{13}$.

From our fiducial sample of 1666 total galaxies (361 quenched) with $M_{*} \geqslant 10^{9.5}~M_{\odot}$ by $z = 0$, 74 (54) reside in a cluster environment at $z = 0$, while 191 (100) are in high-mass groups and 162 (67) in low-mass groups. This leaves 1239 (140) residing in galaxy-scale halos at $z = 0$, of which 999 (104) are primaries while 240 (36) are satellites. We tabulate these results for the quenched population in the right column of Table~\ref{tab:population_environments}.

\begin{deluxetable}{ccccc}[t]
    \tablecaption{Environment statistics for each quenched population at $z = 0$, where primary$/$central galaxies (listed first) are separated from satellite galaxies (listed second in parentheses).\label{tab:population_environments}}
    \tablehead{\colhead{} & \colhead{Inside-out} & \colhead{Outside-in} & \colhead{Ambig.} & \colhead{Total}}
    \startdata
    cluster              &  0  (7) &  0 (39) &  0  (8) &   0 (54)  \\
    high-mass group      &  0 (11) &  0 (68) &  0 (21) &   0 (100) \\
    low-mass group       &  0 (10) &  1 (36) &  0 (20) &   1 (66)  \\
    galaxy-scale (field) & 38 (12) & 30 (11) & 36 (13) & 104 (36)  \\
    \enddata
\end{deluxetable}

To determine the time when a given galaxy was first accreted onto a larger halo, thereby becoming a satellite, we work backward (from $z = 0$) through the main progenitor branch of each quenched galaxy, recording the primary flag at each snapshot. We identify galaxies that have never been a satellite as primary galaxies. For all other cases, to account for minor stochasticity in the primary flags, driven by the ``subhalo switching'' problem \citep[e.g.,][]{rodriguez2015}, where in consecutive snapshots a subhalo is first determined to be a primary, then a satellite, then subsequently a primary again, we identify epochs where a given galaxy is considered a satellite for three consecutive snapshots (generally ${\gtrsim} 470~\text{Myr}$). Using these identifications, we determine the length of time that a galaxy is considered a satellite, and compare this duration with the dynamical time at the beginning of that interval. We adopt the dynamical time scaling relation \citep[e.g.,][]{tinker2010,wetzel2013}
\begin{equation}
    t_{\text{dyn}} = t_{0} \left( 1 + z \right)^{-3/2},
\end{equation}
where $t_{0}$ is set as $2~\text{Gyr}$. We require the galaxy to have been a satellite for at least one dynamical time, and use the earliest such valid occurrence as the time when the galaxy was first accreted onto a more massive structure, making it a satellite. We subsequently refer to this as the satellite time.

We next compare the satellite time with the time when the galaxy first began its primary quenching episode. In Figure~\ref{fig:satellite_time}, we plot the quenching onset time as a function of satellite time for the inside-out population (left panel) and the outside-in population (right panel). In both panels, the points are scaled by their $z = 0$ stellar mass, and galaxies that have always been primaries are plotted in a separate region beyond the current age of the Universe (gray shaded region), though this is for visualization purposes only. Points are color-coded according to their host halo mass at the onset of quenching. Galaxies below the line of equality begin quenching before being accreted, and those above are first accreted and then begin to quench. From Figure~\ref{fig:satellite_time}, it is clear that there are a number of satellites (for both populations) that were accreted many gigayears ago. These satellite galaxies have not been disrupted or have merged into their halo's primary galaxy, as they remain distinct at $z = 0$. Though we might expect that these systems would eventually merge with their primary galaxy over several dynamical times, it has been shown that the merger timescale of satellites depends strongly on the satellite--host mass ratio \citep[e.g.,][]{boylan2008,poulton2020}, where for small mass ratios the merger timescale can be many times longer than the dynamical time.

For the inside-out population in the left panel of Figure~\ref{fig:satellite_time}, nearly half (38$/$78) of the population are primary galaxies at $z = 0$ (far right). Many of the inside-out galaxies began quenching in galaxy-scale halos and show no correlation between their respective times (galaxies below the line of equality). However, for a subpopulation (27) that resided in more massive halos ($M_{\text{h}} \gtrsim 10^{12.3}~M_{\odot}$) when they began quenching, there is some correlation between their satellite time and quenching onset time. For this subpopulation, after being accreted onto a more massive structure, there is an average delay of ${\sim} 1~{\text{Gyr}}$ before quenching begins, indicating that the environment may play a role in promoting quenching for these galaxies.

For the outside-in population (right panel), the large majority (154$/$185) of the population are satellite galaxies, with a subpopulation (31) of primary galaxies (far right). We see that most outside-in quenched galaxies resided in higher-mass halos at quenching onset, with group- and cluster-mass halos being common. Many of the outside-in galaxies in these massive halos show a correlation between their respective times, indicating that the local environment is influencing their evolution through quenching, as expected. For a subpopulation (139) that resided in more massive halos ($M_{\text{h}} \gtrsim 10^{12.3}~M_{\odot}$) at quenching onset, there is an average delay of ${\sim} 1~\text{Gyr}$ after becoming a satellite before quenching began (galaxies above the line of equality). In general, most galaxies that quench as satellites do so in a manner consistent with an outside-in morphological evolutionary signature, while most primary$/$central galaxies quench inside-out. As well, outside-in quenched galaxies are more likely to have a delay before quenching after being accreted, while inside-out quenched galaxies are more likely to have begun quenching prior to becoming a satellite, though this is for inside-out quenched galaxies that resided in galaxy-scale halos at quenching onset. For the inside-out population with $M_{\text{h}} \gtrsim 10^{12.3}~M_{\odot}$, there is a similar delay as for the outside-in population.

As an alternative to considering the different quenched populations per environment, we can instead consider the environments and the populations that reside in them. We summarize the above noted points for $z = 0$ in Table~\ref{tab:population_environments}, where we list the number of primary and satellite galaxies for each quenched population as a function of environment. It is clear that most quenched field galaxies quenched inside-out, whereas satellite galaxies in groups and clusters have a mixture of inside-out and outside-in signatures, though most of the inside-out quenched galaxies began quenching in galaxy-scale halos (see above). Additionally, nearly all $z = 0$ quenched galaxies in groups and clusters are satellites, with only a single central$/$primary quenched galaxy residing in a low-mass group.

\begin{figure}
    \centering
    \includegraphics[width=\columnwidth]{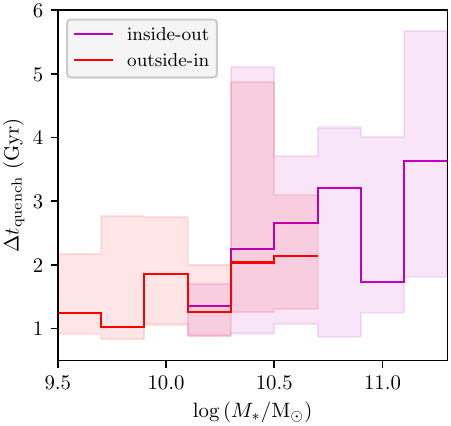}
    \caption{The quenching episode duration (see Section~\ref{subsubsec:quenched}) as a function of stellar mass, divided into the inside-out population (magenta) and outside-in population (red). The inside-out population has higher mass overall, compared to the outside-in population, but shows additional scatter in terms of the quenching duration compared to the outside-in population.}
    \label{fig:quenching_duration}
\end{figure}

We can additionally consider the duration of the quenching episode for each population. In Figure~\ref{fig:quenching_duration}, we show the duration of the quenching episode as a function of stellar mass, grouped by quenched population, with the inside-out population shown in magenta and the outside-in population in red. We use fixed stellar mass bins of $0.2~\text{dex}$ width from $10^{9.5}~M_{\odot}$ to $10^{11.5}~M_{\odot}$ and subsequently require that each stellar mass bin contains a minimum of seven galaxies for reliable statistics. We find that massive ($M_{*} \gtrsim 10^{10.5}~M_{\odot}$) quenched galaxies overwhelmingly quench inside-out, while for low-mass galaxies ($M_{*} \lesssim 10^{10.3}~M_{\odot}$) quenching mostly proceeds outside-in. Within each stellar mass bin, we then compute median quenching episode durations (solid lines) along with ${\pm} 1 \sigma$ values (shaded regions), measured in Gyr. We see that the inside-out population has, in general, longer durations for quenching compared to the outside-in population, but with additional scatter at all masses. In particular, at high masses, quenching durations can last for ${\sim} 3.5~\text{Gyr}$ for the inside-out population, though the average duration is ${\sim} 2.5~\text{Gyr}$. The outside-in population shows less scatter and has a characteristic quenching duration of ${\sim} 1.5~\text{Gyr}$, which has a weak mass dependence.

In summary, though the scatter is larger than the trend seen in Figure~\ref{fig:quenching_duration}, with the morphological metrics alone we can estimate the mean quenching duration on a per-population basis, after classifying galaxies into different morphological evolutionary signatures.

\subsection{Metric evolution per population}

\begin{figure*}[t]
    \centering
    \includegraphics[width=\textwidth]{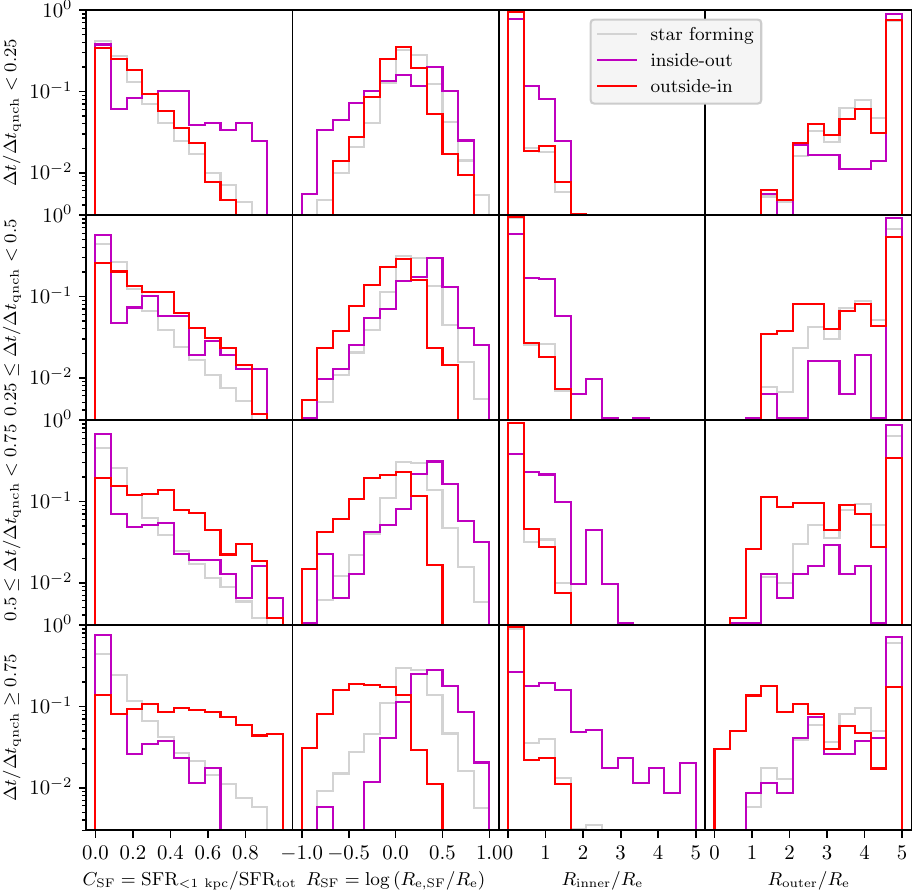}
    \caption{Evolution of the morphological metrics for the entire control star forming sample (gray) as well as the inside-out (magenta) and outside-in (red) quenched populations. Each metric is displayed as a separate column, while each row shows a different epoch through the quenching episode, shown on the \textit{y}-axis. Moving from left to right, the columns show $C_{\text{SF}}$, $R_{\text{SF}}$, $R_{\text{inner}}$, and $R_{\text{outer}}$, respectively. The top row corresponds to $\Delta t = t - t_{\text{onset}} < 0.25 \Delta t_{\text{quench}}$, where $t$ is the time through the quenching episode. The remaining rows similarly contain quartiles of the quenching episode, and moving from top to bottom, the evolution of a given metric is visible. We normalize the distributions so that the \textit{y}-axis in each panel can simply be read as the frequency of that value occurring in the sample.}
    \label{fig:metric_global_evolution}
\end{figure*}

With our quenched sample appropriately classified into various morphological quenching signatures (Section~\ref{subsec:signatures}), we investigate the evolution of the morphological metrics per population. We bin the metrics by evolutionary signature, and consider characteristic epochs through the quenching episode. In Figure~\ref{fig:metric_global_evolution}, we show the distributions of each of the morphological metrics as a separate column, while four characteristic epochs through the quenching episode are displayed as rows. By moving from top to bottom along a column, we see the evolution of a given metric for the populations shown. Given the number of snapshots available for each quenched galaxy along its quenching episode, in combination with the corresponding snapshots for the control star forming galaxies, in total we have 69,493 unique ``galaxies'' from which we consider the evolution of the different populations. In Appendix~\ref{app:ambiguous_histogram_evolution}, we show similar histograms for the ambiguously quenched population.

As mentioned at the end of Section~\ref{subsubsec:quenched}, we additionally define a sample of control star forming galaxies with which to compare the inside-out and outside-in quenched galaxies. These galaxies remain on the star forming main sequence throughout the entirety of a particular quenched galaxy's quenching episode. Given that these galaxies have no quenching episode themselves, we adopt the corresponding timestamps through the quenching episode for the companion quenched galaxy, in order to facilitate comparisons with the ``quenching episode'' of the star forming galaxy. Beginning with the control star forming population (shown in gray), we find no evolution of $C_{\text{SF}}$ and $R_{\text{SF}}$, with $R_{\text{SF}}$ peaking around ${\sim} 0.25$, showing that star formation is generally distributed at larger radii beyond the very central regions. Likewise, we see negligible evolution for the star forming population for $R_{\text{inner}}$ as well as $R_{\text{outer}}$, with both showing no significant cutoff for most galaxies.

Moving next to the inside-out population (shown in magenta), we find this population evolves to low values of $C_{\text{SF}}$, starting from a slightly flatter distribution. We expect this flatter distribution at early times is due to a burst of star formation in the core that is preceding quenching, and this is consistent with how we have defined the onset of quenching, which is based on significant peaks of star formation through the SFH. At late times, there is some similarity between the star forming population and the inside-out population for low values of $C_{\text{SF}}$, though the inside-out population notably cuts off near ${\sim} 0.65$. For $R_{\text{SF}}$, the inside-out population evolves to higher values of $R_{\text{SF}}$, which is expected given the suppression of star formation within the central regions, pushing the subsequent location of the effective radius of the star forming particles to larger radii. This population shows strong evolution of $R_{\text{inner}}$, being most similar to the star forming population early in the quenching episode, but evolving to have a flatter distribution late in the quenching episode. This is expected as $R_{\text{inner}}$ is sensitive to large suppressions of star formation in the inner regions of our quenched galaxies. Finally, when investigating $R_{\text{outer}}$, we find little evolution to lower values for the inside-out population, and they remain largely similar to the star forming population. However, there is a non-negligible subpopulation that has $R_{\text{outer}} < 5$ at all times, indicating that even these quenched galaxies, which have been classified as having an inside-out morphological evolutionary signature, may in fact also experience low levels of outside-in evolution, where this outside-in evolution is weak compared to the strong inside-out evolution.

Finally, for the outside-in population (shown in red), we find an evolution of $C_{\text{SF}}$ in the opposite direction to the inside-out population, evolving to high values of $C_{\text{SF}}$, which is consistent with expectations for this quenching signature, where all the remaining star formation is collected within the central regions. In contrast to the inside-out population, at early times, the outside-in population has a distribution similar to that of the star forming population, indicating that the star formation is more evenly distributed throughout these galaxies prior to quenching. This population also evolves to low values of $R_{\text{SF}}$, and as with $C_{\text{SF}}$, the two quenched populations have $R_{\text{SF}}$ values that evolve in different manners. These low $R_{\text{SF}}$ values reflect how the star formation becomes concentrated in the inner regions, in fact becoming more concentrated than the older stellar populations that trace the distribution of stellar mass, leading to $R_{\text{SF}}$ values that peak at negative values, given the logarithmic definition of $R_{\text{SF}}$. For $R_{\text{inner}}$, we see negligible evolution to higher values, with the distributions remaining similar to those of the star forming population at all times. Finally, for $R_{\text{outer}}$, the outside-in quenched galaxies display strong evolution, with a significant number having $R_{\text{outer}} \approx R_{\text{e}}$ by late times, indicating that the star forming disk has shrunk relative to the stellar disk. As before, we expect such behavior, as $R_{\text{outer}}$ is sensitive to suppression of star formation in the outer regions of a galaxy, exactly like our outside-in signature galaxies.

Given that $R_{\text{inner}}$ and $R_{\text{outer}}$ as described above are quoted in relation to $R_{\text{e}}$ (in addition to $R_{\text{SF}}$), it is important to understand the evolution of $R_{\text{e}}$ across the quenched sample. By considering the quenching episodes, we find that $R_{\text{e}}$ modestly evolves to lower values on average, with an average reduction of ${\sim}$10\% from the onset of quenching until the termination of quenching, i.e. $R_{\text{e, term}} \approx 0.9 R_{\text{e, onset}}$, across the entire quenched sample. When considering the inside-out quenched population alone, we find negligible evolution (${<}$1\%) of $R_{\text{e}}$ on average, while for the outside-in quenched population, we find $R_{\text{e, term}} \approx 0.84 R_{\text{e, onset}}$, indicating that the stellar disk is shrinking more strongly for this population than the inside-out population. These findings suggest that, though the stellar disk is shrinking through the quenching episode for most galaxies, this evolution is mild, and so using the effective radii as a comparison point for the morphological metrics is sensible.

\begin{figure*}[t]
    \centering
    \includegraphics[width=\textwidth]{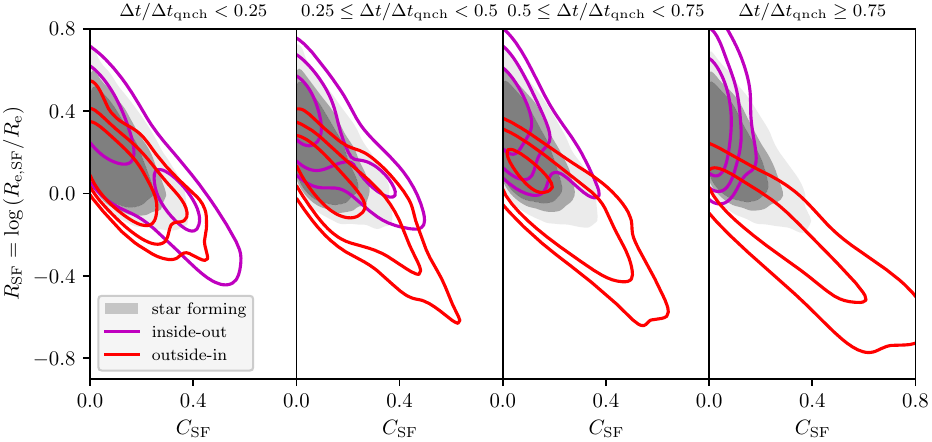}
    \caption{The size of the star forming disk relative to the size of the stellar disk, $R_{\text{SF}}$, as a function of the concentration of star formation in the central kiloparsec relative to the entire stellar disk, $C_{\text{SF}}$. Control star forming galaxies are shown with filled gray contours, where darker colors indicate higher densities. Inside-out quenched galaxies are shown with magenta contours, while outside-in quenched galaxies are shown with red contours. The four panels correspond to quenching episode epoch quartiles, and are the same as in Figure~\ref{fig:metric_global_evolution}.}
    \label{fig:metric_plane_evolution}
\end{figure*}

To demonstrate the robust distinguishing power of these metrics when used in combination, in Figure~\ref{fig:metric_plane_evolution} we show $R_{\text{SF}}$ as a function of $C_{\text{SF}}$ for our inside-out and outside-in quenched galaxies, along with our control sample of normal star forming galaxies. The characteristic epochs through the quenching episode used here are the same as in Figure~\ref{fig:metric_global_evolution}, as is the color scheme. We see that, while the star forming galaxies (gray filled contours) remain relatively fixed from panel to panel, the quenched populations evolve very differently as discussed above. In this space, the inside-out population (magenta contours) begins by largely overlapping with the star forming galaxies, but evolves to the upper left, to low concentrations and high disk ratios. Conversely, the outside-in population (red contours) evolves to the lower right, to high concentrations and low disk ratios. By late times, the populations reside in distinct regions of this parameter space, but with non-negligible overlap with the star forming galaxies. The evolution of these two metrics is not unexpected (see Section~\ref{subsec:signatures} for discussion on an individual galaxy basis), but the amount of ensemble movement throughout the quenching episode is striking.

\subsection{Recovering quenched populations and progress}\label{subsec:ML}

We next determine the ability of machine learning--based algorithms to accurately classify the different populations (star forming, inside-out signature, and outside-in signature) as a function of progress through the quenching episode based on the morphological metrics. We additionally investigate the feasibility of recovering the quenching progress when using machine learning.

\begin{figure*}[t]
    \centering
    \includegraphics[width=\textwidth]{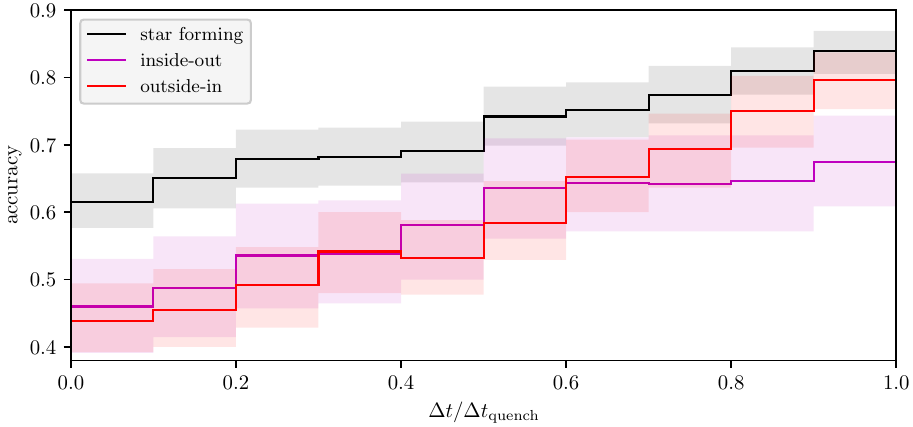}
    \caption{The accuracy of the nearest neighbor--based classification as a function of time (progress) through the quenching episode. The accuracy for the control star forming sample is shown in black, while the inside-out and outside-in quenched populations are shown in magenta and red, respectively. For each population, median values are shown as a thick line, while shaded contours denote the ${\pm} 1\sigma$ range based on 1000 iterations. In these iterations, a randomly selected sample of control star forming galaxies is used, which closely numbers the same amount as the inside-out and outside-in populations.}
    \label{fig:ML_accuracy}
\end{figure*}

We use a \textit{k}-nearest neighbors algorithm \citep[\textit{k}NN; e.g., ][]{fix1951,fix1989,cover1967} available in the package \texttt{scikit-learn} \citep{pedregosa2011}. This algorithm works by finding the \textit{k}-nearest neighbors to a given point and uses these neighbors as training samples to predict the correct label for the point \citep{pedregosa2011}. We use the complete set of morphological metrics and the true population labels (i.e. star forming, inside-out, and outside-in) as the training set samples, and use a 70\%$/$30\% training$/$testing split, where the training$/$testing split is completed such that snapshots from a given galaxy only reside in either the training set or the testing set, but not both, to eliminate bias. Input information related to the progress through the quenching episode is not included, but the progress-binning description below explains how the data are split. We use the three nearest neighbors, and weight the neighbors based on the inverse of their distance. This ensures that closer neighbors of a given point will have a larger influence compared to distant neighbors. Given the distance-based nature of the \textit{k}NN classifier, we additionally standardize$/$feature scale the morphological metrics prior to performing any classification, to ensure that metrics with a larger dynamic range (e.g., $R_{\text{inner}}$ or $R_{\text{outer}}$) do not dominate the distance calculation.

In Figure~\ref{fig:ML_accuracy}, we show the results of applying this machine learning algorithm to our sample, when binned by progress through the quenching episode. For each progress bin, the \textit{k}-nearest neighbors algorithm provides a predicted class based on the morphological metrics, from which we determine the accuracy of the classification by considering the anticipated population and the true population. For each progress bin, we complete 1000 iterations, where in each iteration a randomly selected number of star forming galaxies is used that closely numbers the total number of inside-out and outside-in quenched galaxies, thereby producing a balanced sample. As well, we take the median (solid lines) and ${\pm} 1 \sigma$ range across all iterations (shaded regions). In Figure~\ref{fig:ML_accuracy}, the accuracy of the control star forming population is shown in black, while the inside-out quenched population is shown in magenta, and the outside-in population in red. As above, we are able to discuss the evolution through the ``quenching episode'' for the star forming population, though this is done by using the corresponding time points for the quenched galaxies' quenching episodes. For the star forming population, we find reasonable levels of accuracy (${\sim}$60\%) at early times, with an increase in accuracy values to ${\sim}$70\% by intermediate times. At late times, the accuracy further increases, with values exceeding 80\% for the star forming population. For the inside-out population, at early times ($\Delta t/\Delta t_{\text{quench}} < 0.25$) in the quenching episode, we find lower accuracy values (${\sim} 0.5$). Moving through the quenching episode, we find increasing accuracy values, until at late times ($\Delta t/\Delta t_{\text{quench}} > 0.75$), accuracy values exceed 65\%, though they do not approach the levels of the outside-in population or the star forming population, suggesting that inside-out quenched galaxies are slightly more difficult to classify at late times. This is not completely unexpected, given the evolution of the morphological metrics for the inside-out population as described above in Figure~\ref{fig:metric_global_evolution}. However, we note that the overall rate of increase in accuracy (i.e. slope) for the inside-out population is similar to that of the star forming population. For the outside-in population, we likewise see a clear increase in accuracy with increasing time through the quenching episode, where at early times low accuracy values are apparent (${\sim} 0.45$), while at late times ($\Delta t/\Delta t_{\text{quench}} > 0.75$), the accuracy of the \textit{k}NN classification for the outside-in population can exceed 75\%. This methodology suggests that applying machine learning algorithms to such morphological metrics enables an accurate population prediction, especially in the later stages of quenching. In these later stages, as shown above, the morphological differences between populations is most pronounced.

\begin{figure*}[t]
    \centering
    \includegraphics[width=\textwidth]{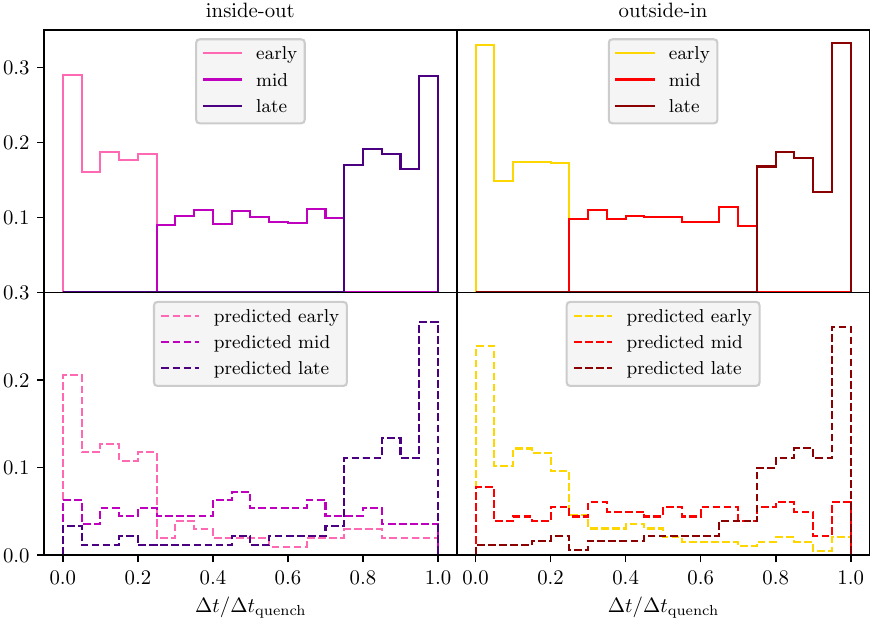}
    \caption{Quenching episode progress for inside-out (left) and outside-in (right) quenched galaxies binned by epoch through the quenching episode, with early-episode galaxies shown in lighter colors, while mid- and late-episode galaxies are shown in progressively darker colors. Top: the true distributions for these galaxies, where we normalize the distributions so that the \textit{y}-axis can simply be read as the frequency of a given value occurring in the sample. The boundary lines that separate the three epochs are easily seen at $\Delta t/\Delta t_{\text{quench}} = 0.25$ and $0.75$. Bottom: the \textit{k}NN--based predictions for the episode progress epochs, after running 1000 iterations and taking the median across all iterations. Colors are as in the top panel, and distributions are similarly normalized.}
    \label{fig:reclassified_quenching_duration}
\end{figure*}

Through testing this classification process, we determined that, by increasing the number of neighbors used, the accuracy of the prediction for the star forming population can be greatly improved. Increasing from three to five or even the 10 nearest neighbors can increase the accuracy beyond 80\% for the majority of the quenching episode, and even beyond 90\% in some cases when considering many neighbors (e.g., $N_{\text{neighbors}} > 25$). However, this increase in the accuracy of the prediction for the star forming population generally produces decreases in the accuracy of the predictions for both quenched populations. At worst, the accuracy values for either population might not exceed 40\%, while the accuracy for the star formers is greater than 90\%. Because of these considerations, we have opted to use three nearest neighbors for the main classification, as three neighbors is possibly optimal for producing the highest accuracy scores for the quenched populations while still providing high values for the star forming galaxies.

We next explore the ability of machine learning to correctly predict the stage at which a quenching galaxy is (i.e. the progress through the quenching episode) using the same \textit{k}NN classifier as above. We consider first the inside-out quenched population alone, then subsequently investigate the outside-in population. As before, we use the three nearest neighbors, weighted by the inverse of the distance to the neighbors, along with the full set of standardized morphological metrics and an unbiased training$/$testing split of 70\%$/$30\%. For the training labels, we use the true progress through the quenching episode, but binned into three distinct epochs. These epochs correspond to early ($\Delta t/\Delta t_{\text{quench}} \leqslant 0.25$), intermediate ($0.25 < \Delta t/\Delta t_{\text{quench}} < 0.75$), and late ($\Delta t/\Delta t_{\text{quench}} \geqslant 0.75$). Using the classifier, we predict the progress epoch and repeat 1000 iterations from which we can take median values. We then follow the same procedure for the outside-in quenched population.

In Figure~\ref{fig:reclassified_quenching_duration}, we show histograms for the quenching progress of the inside-out quenched galaxies (left panels) and the outside-in quenched galaxies (right panels), binned by epoch through the quenching episode. The distributions have been normalized so that the \textit{y}-axis can be read as frequency. In the top panels, we show the true distributions for these galaxies, binned (and color-coded) as described above. In the bottom panels, we show the distributions for the predicted epochs after taking the median across the 1000 iterations. Beginning with the inside-out population and the true distributions, we see noticeable peaks at $\Delta t/\Delta t_{\text{quench}} = 0$ and $\Delta t/\Delta t_{\text{quench}} = 1$ which is a consequence of how we have defined the quenching episode for each quenching galaxy. By construction, every quenching galaxy will have both an onset of quenching as well as a termination of quenching (see near the end of Section~\ref{subsubsec:quenched}). In the bottom panel, we find a larger tail to high quenching episode progress values for the predicted early-episode galaxies, but with a large peak at early values. Similarly, the predicted late-episode galaxies have a quenching episode progress distribution that shows a clear tail to low progress values, mirroring the situation for the predicted early-episode galaxies, with a strong peak at late values. The distribution for the predicted mid-episode galaxies is relatively uniform and lies between the two extreme populations, with tails into both the early and late epochs. This suggests that inside-out galaxies that are at intermediate stages of quenching are more difficult to correctly predict when considering the quenching progress, but that true early-episode galaxies are most commonly classified as early, and similarly true late-episode galaxies are most often classified correctly. Such intermediate-episode galaxies can simultaneously appear as mature early-episode galaxies or as immature late-episode galaxies, producing poor predicted accuracy.

We next consider the outside-in quenched population and largely find a similar situation as for the inside-out population. The \textit{k}-nearest neighbor classifier correctly predicts the true quenching episode epoch most often for early- and late-episode galaxies, as they seem to be distinct morphologically, but intermediate-episode galaxies are more difficult to distinguish. Nonetheless, these findings suggest that machine learning--based algorithms can properly distinguish the progress through the quenching episode for two distinct populations of quenching galaxies, those with an inside-out evolutionary signature, and those with an outside-in evolutionary signature.

\section{Discussion}\label{sec:discussion}

Given the results presented in the current work, simple observationally motivated morphological metrics that are based on the spatial distribution of star formation are capable of being used to identify and distinguish different populations of quenched galaxies within the TNG50 simulation. These quenched populations have different morphological evolutionary signatures, where the two main signatures are consistent with an inside-out evolution and an outside-in evolution. These populations have measurable quenching episodes with differing characteristic durations: that for the inside-out group is ${\sim} 2.5~\text{Gyr}$ on average, but up to ${\sim} 3.5~\text{Gyr}$ for massive galaxies. Meanwhile, the timescale for the outside-in population is shorter and less variable, typically ${\sim} 1.5~\text{Gyr}$. They also exhibit differing characteristic stellar masses, where the outside-in population is generally lower mass than the inside-out population. We have found that these populations are also found (generally) in different environments, where the inside-out population is more likely to be found in the field while the outside-in population is most often found in denser environments (galaxy groups and clusters). An additional difference is that inside-out quenched galaxies are more likely to be the most massive galaxies within their respective dark matter halos, while outside-in galaxies are commonly satellites and begin quenching ${\sim} 1~\text{Gyr}$ after being accreted onto a more massive structure.

Taken together, these results suggest that the nature of galaxy quenching in these simulations is somewhat binary. The most massive galaxies, which reside in the field (and are themselves primary galaxies), experience inside-out quenching, possibly due to the presence of an active galactic nucleus (AGN). This quenching is somewhat extended over relatively long times (compared to their outside-in counterparts) based on the duration of quenching. Conversely, less massive galaxies, which are generally satellites of more massive structures, are first accreted onto those structures and experience a delay before beginning to quench outside-in. This quenching proceeds quickly, taking a few gigayears or less. As well, this outside-in quenching is possibly driven by the dense environment in which these galaxies live, and it could be a consequence of ram pressure stripping.

When considering how these different populations evolve, especially morphologically, the possibility of distinguishing them from each other and from normal star forming galaxies seems feasible. Indeed, by considering the evolution of the observationally motivated morphological metrics, one can make predictions based on those metrics, finding high accuracy for the predicted population at late times when the populations are most distinct. Further, if one considers a given quenched population alone, predictions for the stage of quenching can be made and are largely capable of distinguishing early-stage quenching galaxies from late-stage quenching galaxies.

These results have implications for the field of galaxy evolution, where large samples of quenching galaxies can be determined based on morphology alone, and those samples can further be divided into different quenching evolutionary signatures. As we discuss below, some considerations must be taken before blindly applying these metrics to real galaxies, as the current work has focused on simulated galaxies.

\subsection{Comparison with literature results}

In the context of studies in the literature, we are not presently aware of any work that has completed a directly similar analysis. However, many authors have investigated quenching timescales as a function of galaxy stellar mass for real galaxies \citep[e.g.,][]{wetzel2013,schawinski2014,haines2015,smethurst2015,bremer2018,carnall2018,rowlands2018,phillipps2019,tacchella2022,bravo2023}.

Through a combined analysis of SDSS, GALEX \citep{martin2005}, and Galaxy Zoo \citep{lintott2008} data, \citet{schawinski2014} find that local ($0.02 < z < 0.05$) late-type galaxies ($10^{9} < M_{*}/M_{\odot} < 10^{11.5}$) experience a gradual shutdown of star formation over several gigayears (${\gtrsim} 1~\text{Gyr}$), while local early-type galaxies in a similar mass range experience a very rapid quenching event (${\lesssim} 250~\text{Myr}$) with a simultaneous morphological transformation. \citet{schawinski2014} conclude that the slow quenching of their late-type galaxies is due to a stoppage of the comic gas supply followed by an extended exhaustion of the remaining gas, while the early-type galaxies have gas supplies and reservoirs that are destroyed almost instantaneously, possibly by major mergers. In an analysis that leverages similar SDSS$/$GALEX+Galaxy Zoo data, \citet{smethurst2015} find that, for local to intermediate-redshift galaxies ($0.01 < z < 0.25$; $10^{9} < M_{*}/M_{\odot} < 10^{11.2}$), the quenching timescales are clearly different for smooth (i.e. early-type) and disk-like galaxies. Smooth galaxies are found to undergo rapid quenching (${<} 1~\text{Gyr}$) via major mergers, disk-like galaxies quench over longer periods (${>} 2~\text{Gyr}$) through secular evolution, and an intermediate population quench over intermediate timescales (between $1$ and $2~\text{Gyr}$) via minor mergers and galaxy interactions \citep{smethurst2015}. The results of \citet{smethurst2015} are in qualitative agreement with our results for the quenching timescales (${\sim} 1~\text{Gyr}$ to ${\sim} 2.5~\text{Gyr}$) over a similar mass range ($M_{*} \lesssim 10^{11}~M_{\odot}$), though we do not separate galaxies based on visual morphology (e.g., smooth or disk-like). At higher masses, we find inside-out quenched galaxies can take up to ${\sim} 3.5~\text{Gyr}$ to quench (on median), while \citet{smethurst2015} also find a small percentage of their galaxies can quench over very long timescales (${>} 3~\text{Gyr}$). However, when we compare our results with those of \citet{schawinski2014}, we do not find any galaxies in our sample that quench as quickly as their rapidly quenching galaxies (${\lesssim} 250~\text{Myr}$), with only two galaxies in our sample quenching in less than $500~\text{Myr}$. Even if we account for the perturbations in the duration of our quenching episodes due to different peaks in the star formation history or different definitions of the onset of quenching (see Section~\ref{subsubsec:quenched}), which can be up to ${\sim} 200~\text{Myr}$ when considering the median duration, this would still only influence two galaxies to quenching timescales of ${<} 300~\text{Myr}$. Even so, we find that our results for quenching timescales that take a few gigayears are consistent with the results of \citet{schawinski2014} for their late-type galaxies.

Considering the nature of quenching timescales for local ($z < 0.06$) galaxies in GAMA, \citet{rowlands2018} find a green valley transition timescale of $2.6~\text{Gyr}$ for $M_{*} > 10^{10.6}~M_{\odot}$, which increases with increasing stellar mass. \citet{bremer2018} find that, for intermediate-mass GAMA galaxies ($10^{10.25} < M_{*}/M_{\odot} < 10^{10.75}$) at nearby redshifts ($0.1 < z < 0.2$), the typical green valley traversal timescale is ${\sim}$1--2~Gyr and is independent of environment. Using the same redshift slice but expanding the considered mass range, \citet{phillipps2019} predict that GAMA green valley galaxies have star formation that decreases over 1.5--4~Gyr. As well, \citet{bravo2023} consider quenching timescales and find values ${\sim} 2.5~\text{Gyr}$ for local ($z < 0.06$) low-mass galaxies (${\sim} 10^{9}~M_{\odot}$), but find strong mass dependence for the quenching timescale, where by ${\sim} 10^{11}~M_{\odot}$ the timescale has decreased to $1~\text{Gyr}$. \citet{bravo2023} additionally find an environmental dependence for galaxies less massive than ${\sim} 10^{10}~M_{\odot}$, where satellites have a shorter quenching timescale compared with central galaxies of ${\sim} 0.4~\text{Gyr}$. We find that our results show good agreement with the predictions of \citet{phillipps2019} and also the typical timescale of $2.6~\text{Gyr}$ for high-mass galaxies from \citet{rowlands2018}, in particular the notion that quenching timescale increases with increasing stellar mass. As well, our results are consistent with those of \citet{bremer2018} over their considered mass range. However, when we investigated the environment of the quenched populations, we found that inside-out quenched galaxies are commonly massive galaxies living in the field, which take a long time to quench (${\sim} 2.5~\text{Gyr}$ on average, up to ${\sim} 3.5~\text{Gyr}$ at high masses), while the outside-in population are satellites in dense environments and have shorter quenching durations (${\sim} 1.5~\text{Gyr}$), and so we find a clear environmental trend for the quenching timescale. \citet{bremer2018} finds no environmental dependence for their quenching timescales.

We now discuss studies that consider quenching timescales of simulated galaxies as part of large cosmological simulations \citep[e.g.,][]{trayford2016,hahn2017,nelson2018,rodriguezmontero2019,wright2019,wright2022,park2022,walters2022,wang2023}. \citet{trayford2016} use $10^{9} < M_{*}/M_{\odot} < 10^{11.3}$ galaxies in the Evolution and Assembly of Galaxies and their Environments (EAGLE) simulation \citep{crain2015,schaye2015}. They conclude that most galaxies spend less than $2~\text{Gyr}$ in the green valley by studying intrinsic \textit{u}---\textit{r} colors, and that this timescale is independent of quenching mechanism for satellites and those that host an AGN. Meanwhile, \citet{nelson2018} investigate the formation of the red sequence and the build-up of the color--mass plane using simulated \textit{g}---\textit{r} colors for $10^{9} < M_{*}/M_{\odot} < 10^{12}$ galaxies within TNG100 and TNG300 (larger simulation boxes using the same models as TNG50, but with lower mass resolution) and find a median green valley transition timescale of ${\sim} 1.6~\text{Gyr}$ which decreases as a function of increasing stellar mass (starting at 2--3 Gyr for $M_{*} < 10^{10}~M_{\odot}$ and ending at ${\sim} 1~\text{Gyr}$ for $M_{*} > 10^{11}~M_{\odot}$). They find a similar timescale across TNG100 and TNG300, with little difference between central galaxies and satellites \citep{nelson2018}. In addition, \citet{wright2019} consider two quenching timescales within EAGLE: the green valley traversal timescale on the color--mass diagram, and the timescale for a galaxy to leave the star forming main sequence and settle onto the passive cloud in sSFR--mass space. They find that low-mass centrals ($M_{*} < 10^{9.6}~M_{\odot}$) take longer to quench than satellites of a similar mass (${\sim} 4~\text{Gyr}$ compared to ${\sim} 2~\text{Gyr}$), where the satellite quenching is driven by strong ram pressure stripping and the central quenching is the result of stellar feedback \citep{wright2019}. Higher-mass EAGLE galaxies ($M_{*} > 10^{10.3}~M_{\odot}$) quench more rapidly, ${\lesssim} 2~\text{Gyr}$, for both centrals and satellites, as a consequence of galaxy mergers and AGN feedback. In a result similar to that of \citet{nelson2018}, \citet{wright2019} also find that intermediate-mass galaxies have the longest quenching timescales for centrals and satellites. When we compare our results, we find agreement with \citet{trayford2016} for our outside-in population, where those galaxies have quenching timescales ${<} 2~\text{Gyr}$, and even some of our low-mass inside-out quenched galaxies also have timescales that are consistent with ${<} 2~\text{Gyr}$. For our high-mass inside-out population, we find longer timescales of up to ${\sim} 3.5~\text{Gyr}$. As above, we find a timescale dependence that is linked with quenching mechanism, a result not found by \citet{trayford2016}. Our result of increasing quenching timescale with increasing stellar mass is not shared by \citet{nelson2018} or \citet{wright2019}, who found the opposite relationship, nor is our conclusion of different timescales for centrals$/$satellites when we compare with \citet{nelson2018}, though \citet{wright2019} find a similar result at low mass. Nonetheless, we agree with \citet{nelson2018} on the median quenching timescale when we consider our outside-in galaxies (${\sim} 1.5~\text{Gyr}$ to their ${\sim} 1.6~\text{Gyr}$).

Besides work that investigates quenching timescales, studies in the literature also examine the nature of both inside-out and outside-in quenching in large cosmological simulations, like that of \citet{mcdonough2025}. In their work, \citet{mcdonough2025} study radial profiles of luminosity-weighted ages for low- and high-mass central and satellite galaxies in TNG100. They find that high-mass ($M_{*} > 10^{10.5}~M_{\odot}$) central and satellite galaxies experience inside-out quenching that is driven by AGN feedback, with environmental effects only being apparent at extreme halo masses and local overdensities, while low-mass ($M_{*} < 10^{10}~M_{\odot}$) galaxies experience environmental effects that drive quenching \citep{mcdonough2025}. They additionally find that the environmental quenching proceeds outside-in \citep{mcdonough2025}. We find very good qualitative agreement between our results and those of \citet{mcdonough2025}, in particular with regard to the nature of outside-in quenching for low-mass satellites and inside-out quenching for high-mass centrals.

\subsection{Feasibility with galaxy surveys}

Given the results presented in Figure~\ref{fig:reclassified_quenching_duration}, we find that we can separate early-episode quenching galaxies from quenching galaxies that are late in their quenching episode. This is true for both populations of quenched galaxies investigated in this work. This result suggests that we can estimate the quenching progress for galaxies with these morphological evolutionary signatures. Coupled with the total duration of the quenching episode per population (${\sim} 2~\text{Gyr}$ for inside-out quenched galaxies below $10^{11}~M_{\odot}$, ${\sim} 3.5~\text{Gyr}$ above $10^{11}~M_{\odot}$, and ${\sim} 1.5~\text{Gyr}$ for outside-in quenched galaxies; see Figure~\ref{fig:quenching_duration}), we can make predictions for the elapsed quenching timescale (measured in Gyr) for such galaxies, based only on morphological metrics that trace the evolution of star formation. 

In addition, phase-space diagrams of simulated group and cluster galaxies can provide information on the time since infall for galaxies into a massive halo \citep[e.g.,][]{oman2013,rhee2017,delosrios2021,martinez2023}. With appropriate assumptions about the correlation between time since infall and the onset of quenching, these types of investigations can provide insight into the quenching history of such galaxies \citep[e.g.,][]{lotz2019,pasquali2019,dacunha2022,hough2023}. Our methodology as described above can provide an independent estimate for the onset of quenching. Combining this information with phase-space arguments for time since infall could be a powerful way to investigate when quenching occurs as a function of position on phase-space diagrams using observational data, which is otherwise not possible. This process and the diagnostic information it provides is compelling for future large-scale galaxy surveys, which will have sufficient spatial resolution to map the distribution of star formation for galaxies at various redshifts.

In particular, proposed and upcoming missions like CASTOR \citep{cote2025} and the Ultraviolet Explorer \citep[UVEX;][]{kulkarni2021} aim to image the ultraviolet sky at various resolutions. By combining these observations with other space-based observatories like JWST, NGRST, and Euclid \citep{euclid2025}, all of which will provide longer wavelength coverage to map the distribution of stellar mass, we believe that the morphological metrics described above can be used to distinguish different types of quenching galaxies based on morphological signature. Further, as we have described, these metrics provide great distinguishing power to estimate the duration of quenching for these galaxies, and can place constraints on the quenching progress.

In addition to large photometric galaxy surveys, resolved studies of star formation could prove a compelling testbed for applying our methodology. For example, integral field unit observations can provide high spatial resolution while also collecting spectra. Of the available integral field unit surveys, MaNGA \citep{bundy2015} is the largest, having collected observations for 10$^{\mathrm{4}}$ nearby ($0.01 < z < 0.15$) galaxies \citep{abdurrouf2022}. However, MaNGA splits their sample into a ``primary+'' and secondary sample, where the primary sample covers two-thirds of the available observations and extends to $1.5~R_{\text{e}}$, while the secondary sample is for the remaining third of the sample, with spatial coverage out to $2.5~R_{\text{e}}$ \citep{bundy2015,yan2016,wake2017}. In the current work, we probed the distribution of star formation and stellar mass out to $5~R_{\text{e}}$, and so it is unclear how well our methodology could perform with such a limited extent, if matching to either the MaNGA primary or secondary radial limits. This type of investigation is outside the scope of the present work.

It has been previously suggested that the active galactic nuclei feedback model in TNG is overly strong compared to observations \citep[e.g.,][]{mitchell2020,terrazas2020,donnari2021a,habouzit2022,ma2022,voit2024,wright2024}. This could result in inside-out quenching that is more dramatic than in real data. As well, there is some evidence that points to the mechanism controlling environmental gas removal in TNG being stronger and happening earlier than what observations imply \citep[e.g.,][]{diemer2019,stevens2019,chen2024}. However, other studies have found good agreement between TNG galaxies and real data when investigating environmental gas removal \citep{stevens2021,stevens2023}, indicating that this situation remains unclear. In either event, when considering applying our methodology to samples of real galaxies, care must be taken regarding the strength of inside-out and outside-in quenching. We expect that the morphological metrics that trace sharp truncations of the star forming disk ($R_{\text{inner}}$ and $R_{\text{outer}}$) would be most affected, but these metrics were developed with flexibility regarding their definitions, so they could be readily adapted to real galaxies that have gentler specific star formation radial profiles. As well, these morphological metrics have been found to be less important than $C_{\text{SF}}$ and $R_{\text{SF}}$ for classification purposes (see Appendix~\ref{app:importance}). Beyond the strength of the active galactic nuclei feedback model in TNG, AGN in real galaxies produce emission across the electromagnetic spectrum, including in windows commonly used to trace star formation, such as ultraviolet, H$\alpha$, and infrared emission \citep[e.g.,][]{baldwin1981,veilleux1987,weymann1991,elvis1994,lacy2004,stern2005}. Care must therefore be taken, possibly including the use of molecular gas as a tracer for star formation \citep[e.g.,][]{kennicutt1998a,gao2004,leroy2008,bigiel2008,bolatto2013}.

Of course, observational data for real galaxies does not simply provide star formation rate maps, but one can use ultraviolet and infrared emission to trace the young stellar populations \citep[e.g.,][]{kennicutt1998b,madau2014} or perform spectral energy distribution fitting to obtain star formation rates and stellar masses. In fact, there have already been works analyzing the morphology of stellar mass and star formation rate maps rather than optical light \citep[e.g.,][]{tan2024}. As an intermediate step before applying the current methodology to real galaxies, testing with synthetic imaging (where the underlying true population is known) must be completed \citep[e.g.,][]{smith2018}, and we expect that such testing will reveal additional insights about the feasibility of using morphological metrics like the ones discussed in this work.

\subsection{Limitations and future directions}

When considering the limitations of the current study, we find a few points worth commenting on. The first concerns the definition of the onset of quenching and how this onset is tied to a relative peak in the star formation histories of our sample galaxies. We have found that the distributions of some of our morphological metrics (most obviously $C_{\text{SF}}$, but also $R_{\text{SF}}$ to a lesser degree; see Figure~\ref{fig:metric_global_evolution}) for our inside-out quenched population at early times does not reproduce the distributions for the normal star forming population as expected. As mentioned above, this relative peak in star formation is likely due to a (central) starburst that precedes quenching, leading to the different distributions seen in the top panels of Figure~\ref{fig:metric_global_evolution}. In an ideal sense, we could easily identify when a galaxy begins quenching at any point in the star formation history, but in practice, this becomes difficult to do, which is why we have used such a relative peak to define the quenching onset. As a potential remedy, we could have also considered a few additional snapshots prior to this star formation peak, in order to understand how the morphological metrics evolve and compare with other star forming galaxies. However, this introduces another decision regarding how far back to study, as well as the associated durations of central starbursts. In any event, this type of additional analysis is beyond the scope of the current work.

Additionally, in this work, we have considered all valid quenched galaxies within the simulation, regardless of the redshifts that they began quenching at. As an alternative, we could separate galaxies in the early Universe from those at low redshift and repeat our analyses, given how the morphology of galaxies and their star forming disks evolve through cosmic time \citep[e.g.,][]{shibuya2015,ferreira2023,kartaltepe2023,huertas2024}. However, for our sample of 361 quenched galaxies, we find that 313 began quenching after redshift $z = 1.5$, while 44 began quenching between $1.5 < z < 2.5$, with the remaining galaxies representing a small percentage of our sample (${\sim}$1\%). In addition, we find that ambiguously quenched galaxies make up a larger proportion (than expected) of the sample that began quenching at higher redshifts, and as evidenced above, we do not focus on the ambiguously quenched population in this work. We also find that the median quenching onset redshift is $z \approx 0.73$. With these considerations, we do not expect that separating our sample based on onset redshift should have a large effect on the results presented here, in light of our focus on the inside-out and outside-in quenched populations. Besides, with such a small population that begins quenching at high redshift, statistical comparisons with lower-redshift quenching galaxies might prove difficult.

Regarding future directions, as mentioned above, an appropriate next step is to test this methodology using synthetic imaging or mock observations of these galaxies. In this testing phase, adaptations appropriate for use with observational datasets must be considered, thereby creating observational proxies of the morphological metrics. These proxies must then be compared to understand their performance in recovering the metrics as described in this work. While beyond the scope of the present study, we will next complete this feasibility investigation (Lawlor-Forsyth et al. 2026, in preparation). Beyond these steps, in the event that observational proxies can accurately recover the metrics, then we can look to apply this methodology to datasets of galaxies in the real Universe, to look for samples of inside-out and outside-in quenching galaxies. As discussed above, care must be taken regarding the assumption that the simulated galaxies are representative of the real galaxy population, and additional updates to the definition of the morphological metrics may be necessary. We will aim to use high spatial resolution data that have long integration times, and ideally would use a dataset that includes different environments (e.g., field and dense groups or clusters).

\section{Summary and Conclusion}\label{sec:summary}

In this paper, we have investigated quenched populations and their evolution by considering subhalos present in the highest-resolution run of the IllustrisTNG suite, TNG50. Using the unprecedented resolution of the simulation over a cosmological volume, we first selected a sample of 361 quenched galaxies based on their star formation histories relative to the star formation histories of star forming main-sequence galaxies with a similar stellar mass. Our main findings are as follows:
\begin{itemize}
    \item Four observationally motivated metrics ($C_{\text{SF}}$, $R_{\text{SF}}$, $R_{\text{inner}}$, and $R_{\text{outer}}$), which collectively describe the spatial distribution of star formation, evolve as a galaxy quenches (e.g., Figure~\ref{fig:metric_example_evolution}), with three main modes being prominent: the first where $C_{\text{SF}}$ decreases with time, while $R_{\text{SF}}$ increases with time, and where $R_{\text{inner}}$ strongly evolves, while $R_{\text{outer}}$ is relatively constant. This mode is described as having an inside-out morphological evolutionary signature, and numbers 78 galaxies. The second population has metrics that evolve in the opposite fashion for $C_{\text{SF}}$ and $R_{\text{SF}}$, while $R_{\text{inner}}$ is relatively constant and $R_{\text{outer}}$ evolves strongly, and is described with an outside-in evolutionary signature, numbering 185 galaxies. We find evidence for a third population, which lies between the prior two, and is subsequently considered as having an ambiguously evolving morphological signature, comprising the remaining 98 galaxies. We find no conclusive evidence for an additional class of quenched galaxy where the star formation is uniformly suppressed throughout the galaxy. Such a galaxy would show little evolution in the metrics, with $R_{\text{inner}}$ and $R_{\text{outer}}$ in particular staying essentially constant.
    
    \item Galaxies with an inside-out evolutionary signature are often centrals found in the field, while the environment of outside-in population galaxies tends to be denser regions, like galaxy groups or clusters. Outside-in galaxies are most commonly satellites and experience an average delay of ${\sim} 1~\text{Gyr}$ after being accreted onto a massive structure before beginning to quench (Figure~\ref{fig:satellite_time}). Inside-out and outside-in galaxies are somewhat bimodal in terms of stellar mass, where inside-out quenched galaxies are generally more massive than their outside-in counterparts.
    
    \item Inside-out and outside-in quenched galaxies have quenching timescales that are a function of stellar mass, with low-mass ($M_{*} < 10^{11}~M_{\odot}$) inside-out galaxies taking ${\sim} 2~\text{Gyr}$, while high-mass ($M_{*} > 10^{11}~M_{\odot}$) inside-out galaxies can take up to ${\sim} 3.5~\text{Gyr}$ to quench. Outside-in quenched galaxies have less variability in terms of their quenching timescale, typically taking ${\sim} 1.5~\text{Gyr}$ to quench (Figure~\ref{fig:quenching_duration}).
    
    \item Quenched galaxies evolve in a fundamentally different manner compared to normal star forming galaxies, which do not show significant evolution in their metrics (Figure~\ref{fig:metric_global_evolution}). In parameter space, the inside-out population evolves in a different area than the outside-in population, indicating that this combination of metrics can discriminate between different evolutionary signatures for quenched galaxies, and can also separate such quenched populations from normal star forming galaxies (Figure~\ref{fig:metric_plane_evolution}).
    
    \item \textit{k}-nearest neighbor algorithms can distinguish galaxies based on their location in the metric parameter space, where the classification process accuracy increases as galaxies continue to quench and become morphologically distinct (Figure~\ref{fig:ML_accuracy}). Machine learning can also separate quenched galaxies based on epoch of progress through the quenching episode (Figure~\ref{fig:reclassified_quenching_duration}) for a single population, and combined with estimates for the total duration of the quenching episode, it can provide information on the evolution of these populations and provide testable predictions for observations.
\end{itemize}

Taken holistically, our results suggest that the nature of galaxy quenching is primarily driven by two main mechanisms within the TNG50 simulation: in the first, feedback from active galactic nuclei or strong central stellar feedback suppresses star formation in the central regions of galaxies, and this suppression evolves from the inside out. This mechanism is most commonly found within massive central galaxies, especially those that live within the field. The second mechanism seems to be a consequence of environmental effects, where low-mass satellites of more massive structures experience an environmental gas removal and corresponding suppression of star formation that proceeds from the outside in, consistent with quenching mechanisms such as ram pressure stripping. However, we have also found low levels of outside-in quenching for predominantly inside-out quenched galaxies, indicating that the nature of quenching for simulated galaxies is not wholly binary. Indeed, we have even found high-mass satellites in dense environments that primarily quench from the inside out.

Of course, simulations are an idealized scenario from which to undertake such studies, and this is why we caution against blindly adopting and applying such metrics to observational data without first testing with a control sample with known characteristics. To that end, in future work, we are looking to extend our analysis to synthetic observations of galaxies drawn from TNG50 (Lawlor-Forsyth et al. 2026, in preparation) before applying those findings to observations of galaxies in the real Universe. We are optimistic that the methodology presented in this work will be a powerful tool for upcoming large galaxy surveys that will cover extended areas of the sky with high spatial resolution, enabling the creation of catalogs of galaxies with different evolutionary signatures.

\begin{acknowledgements}
We thank the anonymous referee for providing useful comments that have improved the overall quality of the paper.

The authors would like to acknowledge research grant funding that has enabled this research, including the support of the Natural Sciences and Engineering Research Council of Canada (NSERC) grants RGPIN-2018-03820 and RGPIN-2024-03878 for C.L.F., M.L.B., and C.R.M. S.L.M. acknowledges support from Science and Technology Facilities Council grants ST$/$W000946$/$1 and ST$/$Y000692$/$1. G.H.R. acknowledges the support of National Science Foundation grants AST-2206473 and AST-2308126, and grant 80NSSC21K0641 issued through the NASA Astrophysics Data Analysis Program (ADAP).

The IllustrisTNG simulations were undertaken with compute time awarded by the Gauss Centre for Supercomputing (GCS) under GCS Large-Scale Projects GCS-ILLU and GCS-DWAR on the GCS share of the supercomputer Hazel Hen at the High Performance Computing Center Stuttgart (HLRS), as well as on the machines of the Max Planck Computing and Data Facility (MPCDF) in Garching, Germany.

The authors acknowledge the use of the Canadian Advanced Network for Astronomy Research (CANFAR) Science Platform operated by the Canadian Astronomy Data Center (CADC) and the Digital Research Alliance of Canada (DRAC), with support from the National Research Council of Canada (NRC), the Canadian Space Agency (CSA), CANARIE, and the Canadian Foundation for Innovation (CFI).

This research has made use of the Astrophysics Data System, funded by NASA under Cooperative Agreement 80NSSC25M7105, as well as \texttt{TOPCAT},\footnote{\url{https://www.star.bris.ac.uk/~mbt/topcat}} an interactive graphical viewer and editor for tabular data \citep{taylor2005}, in addition to \texttt{Astropy},\footnote{\url{https://www.astropy.org}} a community-developed core Python package and an ecosystem of tools and resources for astronomy \citep{astropy2013,astropy2018,astropy2022}.
\end{acknowledgements}

\software{\texttt{Astropy} \citep{astropy2013,astropy2018,astropy2022}, \texttt{h5py} \citep{collette2023}, \texttt{Matplotlib} \citep{hunter2007}, \texttt{NumPy} \citep{harris2020}, \texttt{scikit-learn}\footnote{\url{https://scikit-learn.org}} \citep{pedregosa2011}, \texttt{SciPy} \citep{virtanen2020}, \texttt{TOPCAT} \citep{taylor2005}}

\appendix

\section{Example diagnostic figures for an inside-out quenched galaxy}\label{app:io_comprehensive_and_evo}

\begin{figure*}[t]
    \centering
    \includegraphics[width=\textwidth]{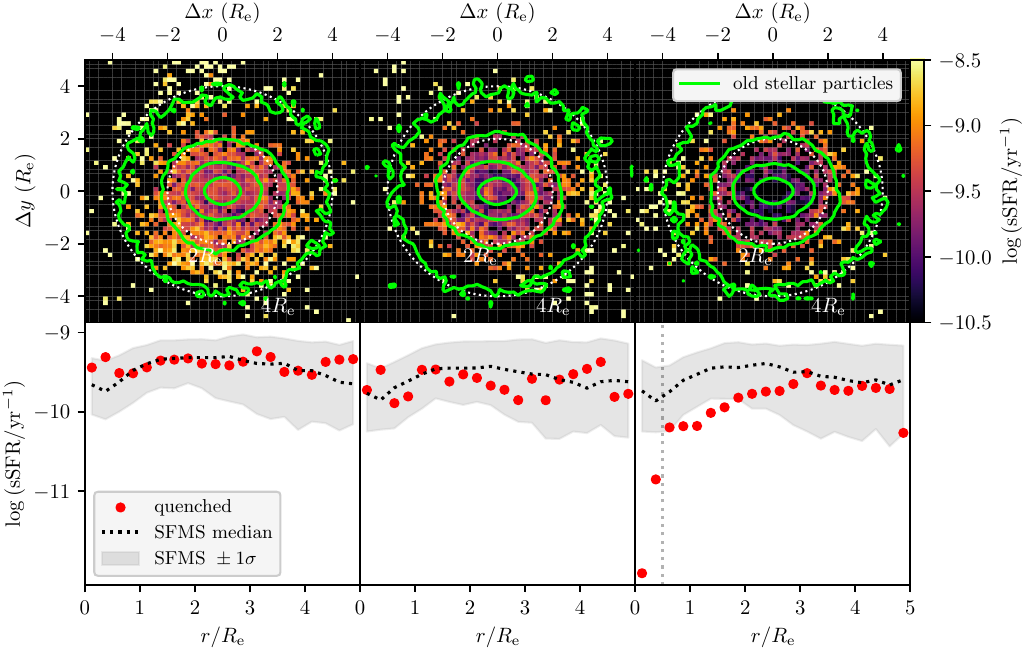}
    \caption{Similar to Figure~\ref{fig:postage_stamps_and_profiles}, but now for an example inside-out quenched galaxy. This galaxy has a $z = 0$ stellar mass of $M_{*} = 10^{10.01}~M_{\odot}$, and the time from the leftmost panels to the middle panels is $336~\text{Myr}$, while the rightmost panels are an additional $306~\text{Myr}$ after the middle panels. The bottom rightmost panel highlights $R_{\text{inner}}$ with a vertical dotted line, where $R_{\text{inner}}/R_{\text{e}} \approx 0.5$. The subhalo ID for this galaxy is 220606 at $z = 0$.}
    \label{fig:postage_stamps_and_profiles_inside-out}
\end{figure*}

Here, we show example postage stamps and radial profiles for an inside-out quenched galaxy in Figure~\ref{fig:postage_stamps_and_profiles_inside-out}, similar to Figure~\ref{fig:postage_stamps_and_profiles} above. The projections in the top panels highlight the lack of evolution of the size of the star forming disk when considering its outer extent, but do show strong evolution from the inside out for the star forming particles. The size of the stellar disk remains nearly unchanged, only growing by ${\sim}$4\%. The radial sSFR profiles in the bottom panels show modest evolution from the leftmost panel to the middle panel, but show very strong evolution from the middle to the rightmost panel. In this case, the rightmost panel highlights $R_{\text{inner}}$, as there is a clear discontinuity in the sSFR profile for this quenched galaxy, where $R_{\text{inner}}/R_{\text{e}} \approx 0.5$.

\begin{figure}
    \centering
    \includegraphics[width=\columnwidth]{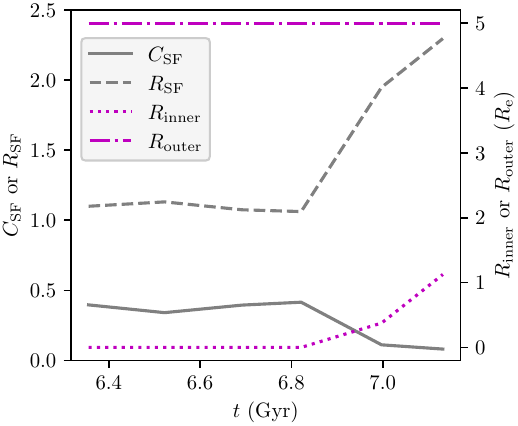}
    \caption{Similar to Figure~\ref{fig:metric_example_evolution}, but now for the same example inside-out quenched galaxy as is shown above in Figure~\ref{fig:postage_stamps_and_profiles_inside-out}. As above, the left \textit{y}-axis records the values for $C_{\text{SF}}$ and $R_{\text{SF}}$, shown in gray solid and dashed lines, respectively. The right \textit{y}-axis records the values for $R_{\text{inner}}/R_{\text{e}}$ (dotted magenta line) and $R_{\text{outer}}/R_{\text{e}}$ (dotted--dashed magenta line). The extent of the \textit{x}-axis reflects the primary quenching episode for this galaxy, which lasts for ${<} 1~\text{Gyr}$.}
    \label{fig:metric_example_evolution_inside-out}
\end{figure}

In addition, we show the evolution of the morphological metrics for the same inside-out quenched galaxy in Figure~\ref{fig:metric_example_evolution_inside-out}, similar to Figure~\ref{fig:metric_example_evolution} above. This galaxy's morphological evolution is consistent with an inside-out signature, seen through the increase of $R_{\text{SF}}$ and $R_{\text{inner}}$, along with $C_{\text{SF}}$ modestly decreasing. This galaxy shows no evolution for $R_{\text{outer}}$, a feature commonly noticed for inside-out quenched galaxies in our sample.

\section{Morphological metrics using 3D positional information}\label{app:3d_metric_comparison}

\begin{figure*}
    \centering
    \includegraphics[width=\textwidth]{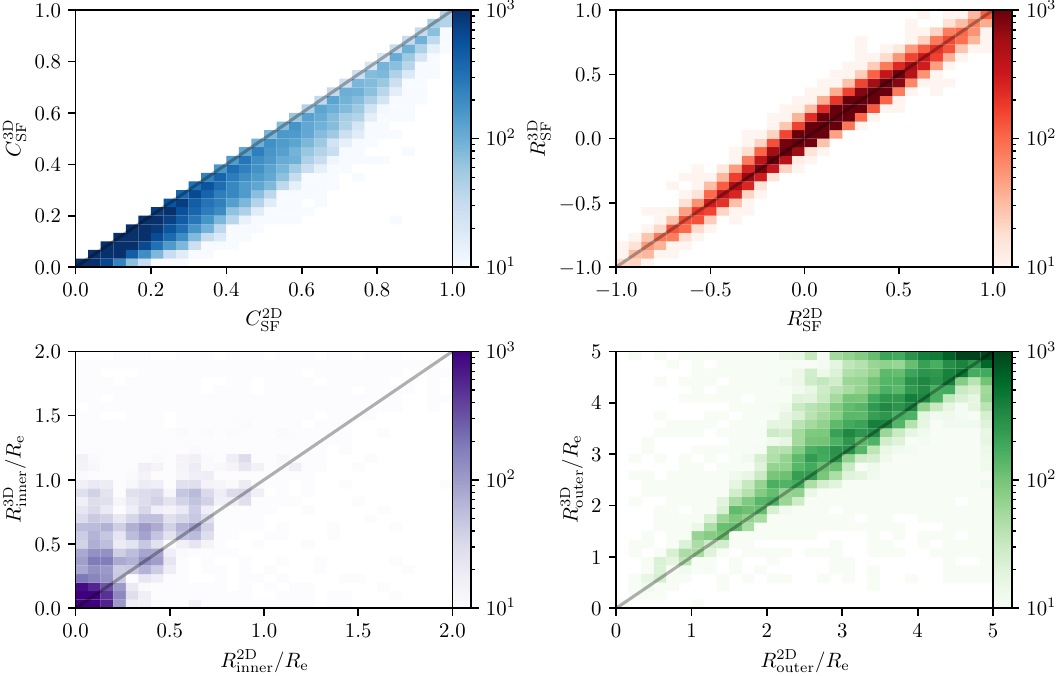}
    \caption{Comparisons between the morphological metrics using the default versions (\textit{x}-axis, where 2D positional information is used) and versions using full 3D positional information (\textit{y}-axis). For all panels, we use 2D histograms to show the density of points, described with the color scale. For the inner truncation radius, we zoom into the region of interest for clarity.}
    \label{fig:metric_comparisons}
\end{figure*}

Here, we present comparisons between the morphological metrics in Figure~\ref{fig:metric_comparisons} for the full sample of quenched galaxies and control star forming galaxies. As mentioned above, our full sample comprises 69,493 unique galaxy+snapshot pairs, which we collectively consider unique galaxies, and which we plot in each panel. We compare the default version of the metric (plotted on the \textit{x}-axis, where 2D positional information was used to compute distances to the stellar particles) to versions using the full 3D positional information to compute distances (\textit{y}-axis). In all panels, we employ a logarithmic normalization given the large number of data points per plot, and also display a line of equality in gray.

For the concentration of star formation, $C_{\text{SF}}$, we find that using the full 3D positional information produces concentration values that are always less than the default 2D versions. This behavior is not unexpected, as the additional information provided in the third dimension is reflected in locations that are more accurate for the stellar particles, and given how the concentration metric is defined, this results in concentration values that are likewise more accurate. We note that the colormap used in this panel creates a false sense of systematic offset, but given the logarithmic normalization that is being used, histogram counts away from the line of equality quickly fall off, with typical values being ${\lesssim} 100$. In general, we find good agreement between the 2D versions of $C_{\text{SF}}$ and the 3D versions.

For the disk size ratio, $R_{\text{SF}}$, we find excellent agreement between the 2D versions and the 3D versions, with small scatter around the line of equality. Moving away from the line of equality, we quickly find histogram bins that contain very few galaxies or no galaxies at all. For the inner truncation radius, $R_{\text{inner}}$, we see that the 3D versions can have modest scatter, especially to slightly higher values, from their 2D counterparts at low $R_{\text{inner}}$ values. Beyond ${\sim} 1.5~R_{\text{e}}$, there is generally very low levels of scatter around the 2D versions, extending to $2~R_{\text{e}}$ and beyond. In general, there exists good agreement between 2D versions and 3D versions of $R_{\text{inner}}$, especially at very low $R_{\text{inner}}$ values (${\sim} 0~R_{\text{e}}$).

Finally, for the outer truncation radius, $R_{\text{outer}}$, we find that 3D versions lie along the line of equality or show modest enhancement compared to their 2D counterparts, below ${\sim} 4~R_{\text{e}}$. Moving above ${\sim} 4~R_{\text{e}}$, 2D versions can take a slightly broader range of values at a fixed 3D version value, with this trend becoming most noticeable when 3D $R_{\text{outer}} \approx 5~R_{\text{e}}$. Similar to the other metrics previously discussed, in general we find very good agreement between the 2D versions of $R_{\text{outer}}$ and the 3D versions, and overall we find agreement for all metrics that ranges from good to excellent. We do not expect that using the 3D versions of our metrics should significantly change the results presented, with only minor effects anticipated, especially given the investigation presented in Appendix~\ref{app:importance} (see below).

\section{Morphological metric evolution plots for the ambiguously quenched class}\label{app:ambiguous_histogram_evolution}

\begin{figure*}
    \centering
    \includegraphics[width=\textwidth]{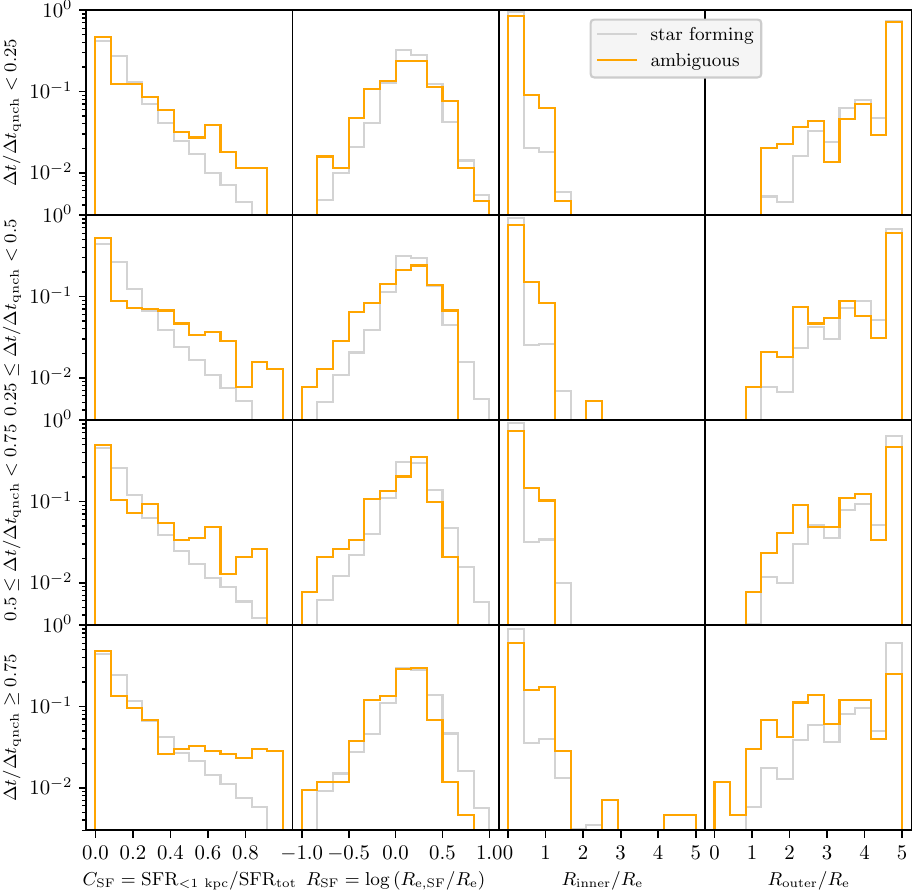}
    \caption{Morphological metric evolution for the ambiguously quenched class of galaxies, similar to Figure~\ref{fig:metric_global_evolution}.}
    \label{fig:metric_global_evolution_amb}
\end{figure*}

Here, we present the evolution of the morphological metrics for the ambiguously quenched class in Figure~\ref{fig:metric_global_evolution_amb}, along with the star forming population for reference. By comparing with Figure~\ref{fig:metric_global_evolution}, which shows the evolution of the metrics for normal star forming galaxies, the galaxies that have an inside-out signature, and the galaxies that have an outside-in signature, we can see that, in most cases, the ambiguous class resides in an intermediary space. Specifically, when comparing $C_{\text{SF}}$ at early times, the ambiguous class is between the inside-out and outside-in classes, and at later times, it remains between these two classes. For $R_{\text{SF}}$, the ambiguous class most closely resembles the star forming class, with little evolution, but with a slight shift to lower values. At late times, the resemblance to the star forming population remains. For $R_{\text{inner}}$, the ambiguous class is most similar to the inside-out class at early times, but at late times, it has evolved to sit between the inside-out and star forming classes. Finally, for $R_{\text{outer}}$, at early times, the ambiguous class resembles the star forming class, but at late times, it evolves in a similar fashion to the outside-in class, though to a lesser extent. Given the above, it is clear that the ambiguously quenched galaxies do reside in an intermediary parameter space between the inside-out and outside-in quenched populations and are therefore properly classified as such.

\section{Relative importance of the morphological metrics}\label{app:importance}

Here, we discuss the relative importance of the morphological metrics when compared to each other for classification and prediction purposes, using a random forest classifier algorithm \citep{breiman2001}. This algorithm is available in the package \texttt{scikit-learn} \citep{pedregosa2011} and uses an ensemble of randomized decision trees, where each tree is created from a sample of the training set drawn using bootstrap sampling (sampling with replacement). The algorithm then uses averaging to control for overfitting and to improve the accuracy of predictions \citep{pedregosa2011}. As in Section~\ref{subsec:ML}, we bin the progress through the quenching episode into deciles, and again randomly select the same number of star forming galaxies as quenched galaxies for each progress bin, using 1000 iterations. We use 100 randomized decision trees per forest.

\begin{figure*}
    \centering
    \includegraphics[width=\textwidth]{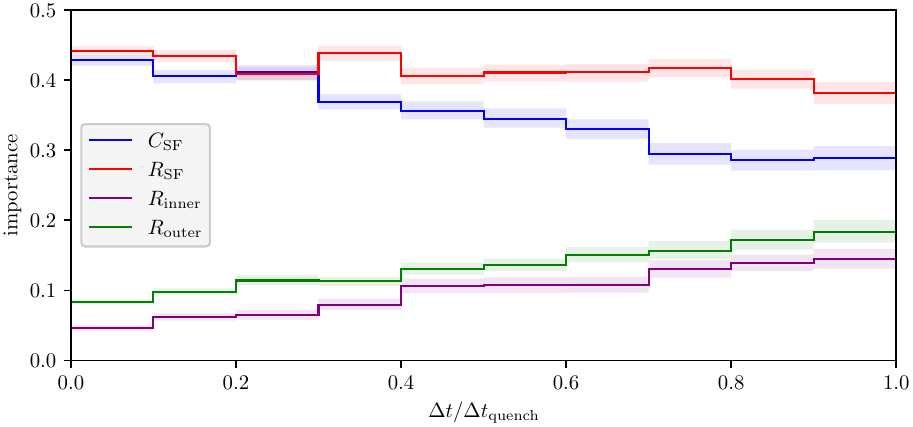}
    \caption{The importance (mean decrease in impurity, or Gini impurity) of the morphological metrics for accurate classification as a function of progress (time) through the quenching episode. For each metric, $C_{\text{SF}}$ (blue), $R_{\text{SF}}$ (red), $R_{\text{inner}}$ (purple), and $R_{\text{outer}}$ (green), we show median values across 1000 iterations as solid lines, with the corresponding shaded contours denoting the ${\pm} 1 \sigma$ range. For each iteration, a randomly selected sample of star forming galaxies is chosen to closely mirror the same number of quenched galaxies (inside-out + outside-in).}
    \label{fig:random_forest}
\end{figure*}

In Figure~\ref{fig:random_forest}, we show the results of this investigation. We find that the ratio between the size of the star forming disk and the stellar disk, $R_{\text{SF}}$ (red), is the most important metric for accurately predicting the true population (see Figure~\ref{fig:ML_accuracy}) for all epochs through the quenching episode. We note the relative decrease in importance at late times, but this is compensated with increases in other metrics, as discussed below. The concentration of star formation, $C_{\text{SF}}$ (blue), is the second most important metric, especially at early times, but decreases in importance through the quenching episode, notably at very late times. The outer truncation radius, $R_{\text{outer}}$ (green), is next most important, with importance values near 0.15 for much of the episode, with a steady increase in importance through the quenching episode clearly visible. This suggests that, at later times, $R_{\text{outer}}$ has more distinguishing power (compared to earlier times) for correctly predicting the true population of a given galaxy. Finally, we find that $R_{\text{inner}}$ has the lowest importance in general, but that the importance gradually increases with increasing time, similar to $R_{\text{outer}}$, with importance ${\sim}$0.15 at very late times. This indicates that the characteristic radius that traces sharp truncations in the inner part of the star forming disk has more distinguishing power once galaxies have had sufficient time to develop a suppression of star formation in their central regions.

Taken together, these results indicate that no single morphological metric, or even two morphological metrics used simultaneously, would be able to properly classify and predict the true population for a collection of star forming galaxies and inside-out and outside-in quenched galaxies. However, one could argue whether it is necessary to include $R_{\text{inner}}$ and $R_{\text{outer}}$ in the above analysis, given their low (${<}$0.1) importance values at very early times. Regardless, this consideration is beyond the scope of the present work.

\bibliography{references}{}
\bibliographystyle{aasjournal}

\end{document}